\pdfoutput=1
\documentclass{PoS}
\usepackage[style=numeric,sorting=none]{biblatex}
\DeclareSourcemap{
  \maps[datatype=bibtex,overwrite=true]{
    \map{
      \step[fieldsource=authortype, final=true]
      \step[fieldset=userb, origfieldval, final=true]
    }
    \map{
      \step[fieldsource=collaboration, final=true]
      \step[fieldset=usera, origfieldval, final=true]
    }
  }
}
\renewbibmacro*{author}{%
    {\printnames{author}%
     \iffieldundef{userb}
       {}
       { on behalf of the \printfield{userb}}
     \iffieldundef{usera}
       {}
       { (\printfield{usera})}
     }
   }

\usepackage{cancel}
\usepackage{slashed}
\usepackage[T1]{fontenc}
\usepackage{textcomp}
\usepackage{lmodern}
\usepackage{rotfloat}
\usepackage{graphicx}
\graphicspath{{./FIGS/}}
\usepackage{color}

\makeatletter
\@ifundefined{texorpdfstring}
{\newcommand\texorpdfstring[2]{#1}}{}
\@ifundefined{unichar}
{\newcommand\myunichardef[3]{\expandafter\providecommand\csname text#1\endcsname
                            {#2}}}
{\newcommand\myunichardef[3]{\expandafter\providecommand\csname text#1\endcsname
                            {\unichar{"#3}}}}
\makeatother
\myunichardef{composetilde}{\string~}{0303}
\myunichardef{composemacron}{bar}{0305}
\myunichardef{subscriptthree}{\string_3}{2083}

\newbox\dpbox
\setbox\dpbox\hbox{}
\dp\dpbox=3\jot

\definecolor{lightgreen}{rgb}{.85,.99,.85}
\definecolor{darkgreen}{rgb}{0,.7,0}
\definecolor{orange}{rgb}{1.0,.6,0}

\newcommand{\rC}{C}
\newcommand{\gA}{A}
\newcommand{\oP}{P}

\newcommand{\bda}{\begin{\displaymath}\begin{array}{rl}}
\newcommand{\eda}{\end{array}\end{displaymath}}
\newcommand{\be}{\begin{equation}}
\newcommand{\ee}{\end{equation}}
\newcommand{\bdm}{\begin{displaymath}}
\newcommand{\edm}{\end{displaymath}}
\newcommand{\bea}{\begin{eqnarray}}
\newcommand{\eea}{\end{eqnarray}}

\newcommand{\off}[1]{{}}

\newcommand{\bi}{\begin{itemize}}
\newcommand{\ei}{\end{itemize}}

\newcommand{\beq}{\begin{equation}}
\newcommand{\eeq}{\end{equation}}

\newcommand{\msbar}{{\overline{{\rm MS}}}}

\def\mev{{\rm MeV}}
\def\gev{{\rm GeV}}
\def\tev{{\rm TeV}}

\newcommand{\bd}{\begin{displaymath}}
\newcommand{\ed}{\end{displaymath}}

\newcommand{\figurebox}[2]{\fbox{\vbox to#2in{\hbox to #1in{\hfil}\vfil}}}
\newcommand{\Nf}{N_{\hspace{-0.08 em} f}}

\definecolor{Gray}{rgb}{0.5,0.5,0.5}
\definecolor{Black}{rgb}{0.0,0.0,0.0}

\def\good{\makebox[1em]{\centering{\mbox{\color{green}$\bigstar$}}}}
\def\bad{\makebox[1em]{\centering\color{red}\tiny$\blacksquare$}}
\def\soso{\makebox[1em]{\centering{\mbox{\raisebox{-0.5mm}{\color{green}\Large$\circ$}}}}}

\def\u#1{{U\left(#1\right)}}
\def\d#1#2{d\mskip 1.5mu^{#1}\mkern-1mu{#2}\,}
\def\D#1#2{{\d#1{#2} \over (2\pi)^{#1}}\,}

\def\ev{\mathrm{e\kern-0.1em V}}
\def\kev{\mathrm{ke\kern-0.1em V}}
\def\mev{\mathrm{Me\kern-0.1em V}}
\def\gev{\mathrm{Ge\kern-0.1em V}}
\def\tev{\mathrm{Te\kern-0.1em V}}

\def\n#1e#2n{{#1}\times 10^{#2}}

\def\bea{\begin{eqnarray}}
\def\eea{\end{eqnarray}}

\def\ods2{\mathcal{O}_{\Delta S=2}}
\def\zds2{Z_{\Delta S=2}}

\def\msbar{{\overline{\mathrm{MS}}}}

\def\spose#1{\hbox to 0pt{#1\hss}}
\def\ltapprox{\mathrel{\spose{\lower 3pt\hbox{$\mathchar"218$}}
 \raise 2.0pt\hbox{$\mathchar"13C$}}}
\def\gtapprox{\mathrel{\spose{\lower 3pt\hbox{$\mathchar"218$}}
 \raise 2.0pt\hbox{$\mathchar"13E$}}}
\def\inapprox{\mathrel{\spose{\lower 3pt\hbox{$\mathchar"218$}}
 \raise 2.0pt\hbox{$\mathchar"232$}}}

\makeatletter
\def\slash#1{{\mathpalette\c@ncel{#1}}} 
\def\big#1{{\hbox{$\left#1\vbox to1.012\ht\strutbox{}\right.\n@space$}}}
\def\Big#1{{\hbox{$\left#1\vbox to1.369\ht\strutbox{}\right.\n@space$}}}
\def\bigg#1{{\hbox{$\left#1\vbox to1.726\ht\strutbox{}\right.\n@space$}}}
\def\Bigg#1{{\hbox{$\left#1\vbox
to2.083\ht\strutbox{}\right.\n@space$}}}
\makeatother

\def\spose#1{\hbox to 0pt{#1\hss}}
\def\ltapprox{\mathrel{\spose{\lower 3pt\hbox{$\mathchar"218$}}
\raise 2.0pt\hbox{$\mathchar"13C$}}}
\def\gtapprox{\mathrel{\spose{\lower 3pt\hbox{$\mathchar"218$}}
\raise 2.0pt\hbox{$\mathchar"13E$}}}
\def\inapprox{\mathrel{\spose{\lower 3pt\hbox{$\mathchar"218$}}
\raise 2.0pt\hbox{$\mathchar"232$}}}

\makeatletter
\def\Hy@colorlink#1{}
\let\Hy@endcolorlink\relax
\makeatother
\usepackage[hyperfootnotes=false]{hyperref}

\title{Recent results of nucleon structure \& matrix element calculations}

\ShortTitle{Nucleon Structure and Matrix Elements}

\author{\speaker{Tanmoy Bhattacharya}, Rajan Gupta, and Boram Yoon\\
        Los Alamos National Laboratory, Los Alamos, NM 87545, USA\\
        E-mail: \email{tanmoy@lanl.gov}, \email{rajan@lanl.gov}, \email{boram@lanl.gov}}

      \abstract{A review of recent lattice calculations of nucleon
        structure and matrix elements of operators in nucleons is
        presented. It primarily covers developments in the
        calculation of the matrix elements of the scalar, tensor,
        pseudo-scalar, axial-vector and vector operators in the ground
        state of neutrons and protons in the isospin symmetric
        limit. Some preliminary calculations of the electric dipole
        moment, the gravitational moments and stress-energy
        distribution, and the magnetic polarizability are briefly
        described.}

\FullConference{37th International Symposium on Lattice Field Theory - Lattice2019\\
		16-22 June 2019\\
		Wuhan, China}

\begin{document}

\section{Introduction}
This review will only cover the matrix elements of local, parity
conserving, dimension-3, quark bilinears operators in the ground state
of neutrons and protons in the isospin symmetric ($m_u=m_d$) limit. In
particular, I will not discuss products or commutators of
currents~\cite{Zimmermann.this, Liang.this, Portelli.this}, contribution to the electric
dipole moments from the QCD \(\Theta\)-term~\cite{Ohki.this}, PDFs and
their moments~\cite{Zhao.this,Karthik.this}, matrix elements within other
baryon states~\cite{Weishaupl.this}, or decays~\cite{Aoki.this}, which are
discussed by other speakers in these proceedings.

The calculational methodology of isovector matrix elements of nucleons
is mature, however, control over all the systematics and assigning a
reliable estimate of uncertainty to each needs to be reviewed. As
discussed below, the errors in the scalar matrix elements are still
large. There are new publications showing excited states with small mass gap
contribute to isoscalar matrix elements at small nonzero momentum,
{\it i.e.\hbox{},} in the form factors, consequently the systematic
error due to the excited state contamination may be
underestimated. There has been no significant development in
the methodology for reducing the systematics due to the
renormalization factor, finite volume, discretization errors and
scale-setting, and the chiral extrapolation to the light quark
mass. Overall, the first part of my talk on nucleon charges has
significant overlap with the recent FLAG 2019
report~\cite{Aoki:2019cca} that provides a status report up to the beginning of 2019, and the reader is referred to it.

\section{Isovector \texorpdfstring{\(g_A\)}{g\string_A}}
\begin{figure}
\begin{minipage}{0.5\linewidth}
  \tiny
  \setlength{\tabcolsep}{0pt}
\begin{tabular*}{\textwidth}[l]{l @{\extracolsep{\fill}} r lllllll l }
Collaboration & $\Nf$ & 
\hspace{1em}\begin{rotate}{90}{pub.}\end{rotate}\hspace{-1em} &
\hspace{1em}\begin{rotate}{90}{cont.}\end{rotate}\hspace{-1em} &
\hspace{1em}\begin{rotate}{90}{chiral}\end{rotate}\hspace{-1em}&
\hspace{1em}\begin{rotate}{90}{vol.}\end{rotate}\hspace{-1em}&
\hspace{1em}\begin{rotate}{90}{ren.}\end{rotate}\hspace{-1em}  &
\hspace{1em}\begin{rotate}{90}{states}\end{rotate}\hspace{-1em}  &
$g^{u-d}_A$\\
\hline
\hline
%
PNDME 18$^a$ &  2+1+1 & \gA & \good$^\ddag$ & \good & \good & \good & \good & 1.218(25)(30) \\[0ex]
CalLat 18 &  2+1+1 & \gA & \soso & \good & \good & \good & \good & 1.271(10)(7) \\[0ex]
CalLat 17 &  2+1+1 & \oP & \soso & \good & \good & \good & \good & 1.278(21)(26) \\[0ex]
PNDME 16$^a$ &  2+1+1 & \gA & \soso$^\ddag$ & \good & \good & \good & \good & 1.195(33)(20) \\[0ex]
\hline
Mainz 18 &  2+1 & \rC & \good & \soso & \good & \good & \good & 1.251(24) \\[0ex]
PACS 18 &  2+1 & \gA & \bad & \bad & \good & \good & \bad & 1.163(75)(14) \\[0ex]
$\chi$QCD 18 &  2+1 & \gA & \soso & \good & \good & \good & \good & 1.254(16)(30)$^\$$ \\[0ex]
JLQCD 18 &  2+1 & \gA & \bad & \soso & \soso & \good & \good & 1.123(28)(29)(90) \\[0ex]
LHPC 12A$^b$ &  2+1 & \gA & \bad$^\ddag$ & \good & \good & \good & \good & 0.97(8) \\[0ex]
LHPC 10 &  2+1 & \gA & \bad & \soso & \bad & \good & \bad & 1.21(17) \\[0ex]
RBC/UKQCD 09B &  2+1 & \gA & \bad & \bad & \soso & \good & \bad & 1.19(6)(4) \\[0ex]
RBC/UKQCD 08B &  2+1 & \gA & \bad & \bad & \soso & \good & \bad & 1.20(6)(4) \\[0ex]
LHPC 05 &  2+1 & \gA & \bad & \bad & \good & \good & \bad & 1.226(84) \\[0ex]
\hline
Mainz 17 &  2 & \gA & \good & \good & \good & \good & \soso & 1.278(68)($^{+0}_{-0.087}$) \\[0ex]
ETM 17B &  2 & \gA & \bad & \soso & \soso & \good & \good & 1.212(33)(22) \\[0ex]
ETM 15D &  2 & \gA & \bad & \soso & \soso & \good & \good & 1.242(57) \\[0ex]
RQCD 14 &  2 & \gA & \soso & \good & \good & \good & \bad & 1.280(44)(46) \\[0ex]
QCDSF 13 &  2 & \gA & \soso & \good & \bad & \good & \bad & 1.29(5)(3) \\[0ex]
Mainz 12 &  2 & \gA & \good & \soso & \soso & \good & \soso & 1.233(63)($^{+0.035}_{-0.060}$) \\[0ex]
RBC 08 &  2 & \gA & \bad & \bad & \bad & \good & \bad & 1.23(12) \\[0ex]
QCDSF 06 &  2 & \gA & \soso & \bad & \bad & \good & \bad & 1.31(9)(7) \\[0ex]
\hline
\hline
\end{tabular*}
{\tiny
$^a$Tree-level tadpole-improved.
$^b$Tree-level improved.
$^\ddag$Not fully O(a) improved.
$^\$$Also has partially quenched analysis.
}
\end{minipage}
\begin{minipage}{0.5\linewidth}
  \vspace*{-\baselineskip}
  \hbox to 0pt{\includegraphics[width=\linewidth,height=16\baselineskip]{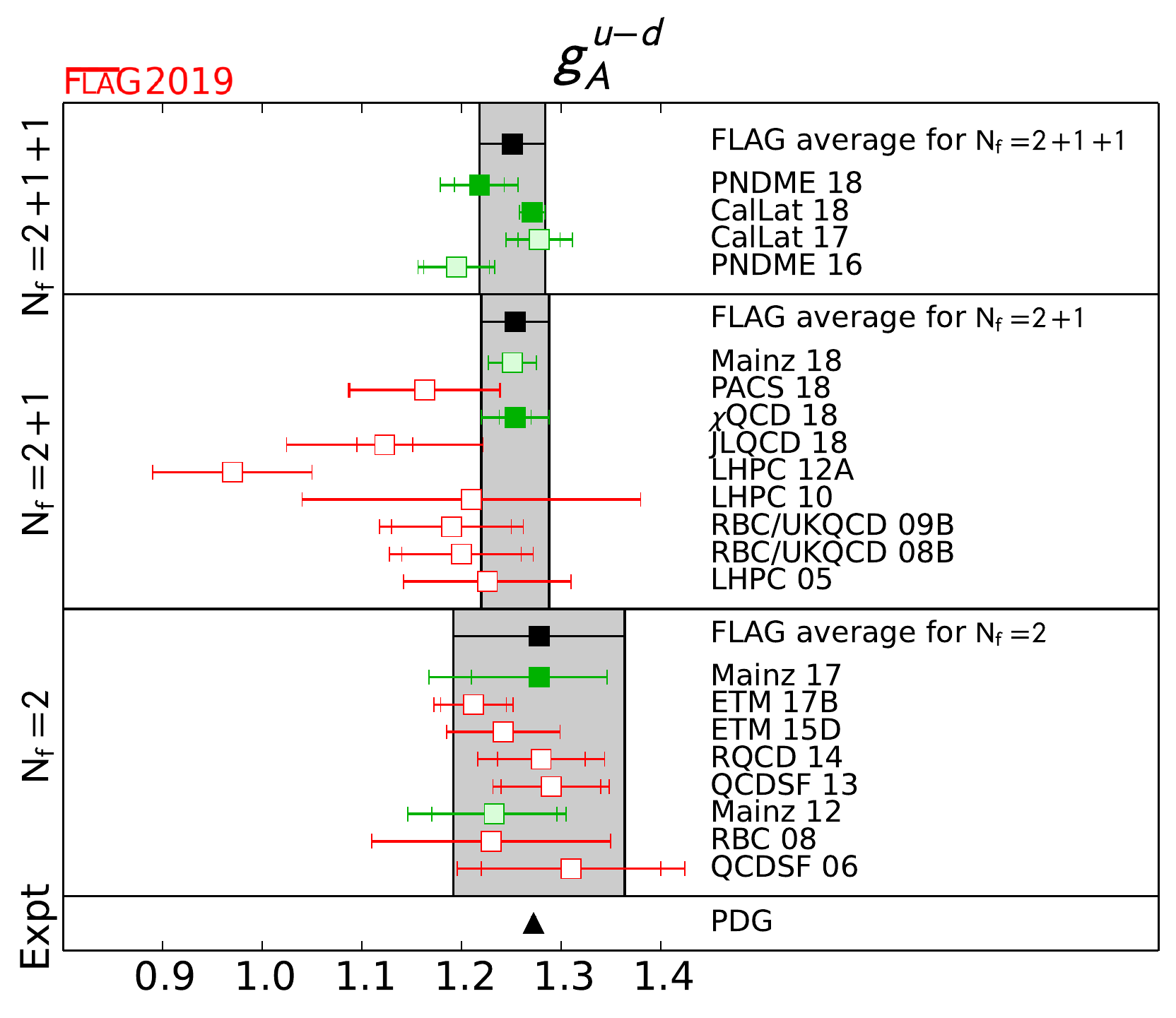}\hss}
\end{minipage}
 %
  \caption{Compilation of the world data on isovector \texorpdfstring{\(g_A\)}{g\string_A} presented in the FLAG~\protect\cite{Aoki:2019cca} Review.}
  \label{fig:gAFLAG}
\end{figure}
\begin{figure}
\begin{center}
\leavevmode
 \includegraphics[width=0.99\linewidth]{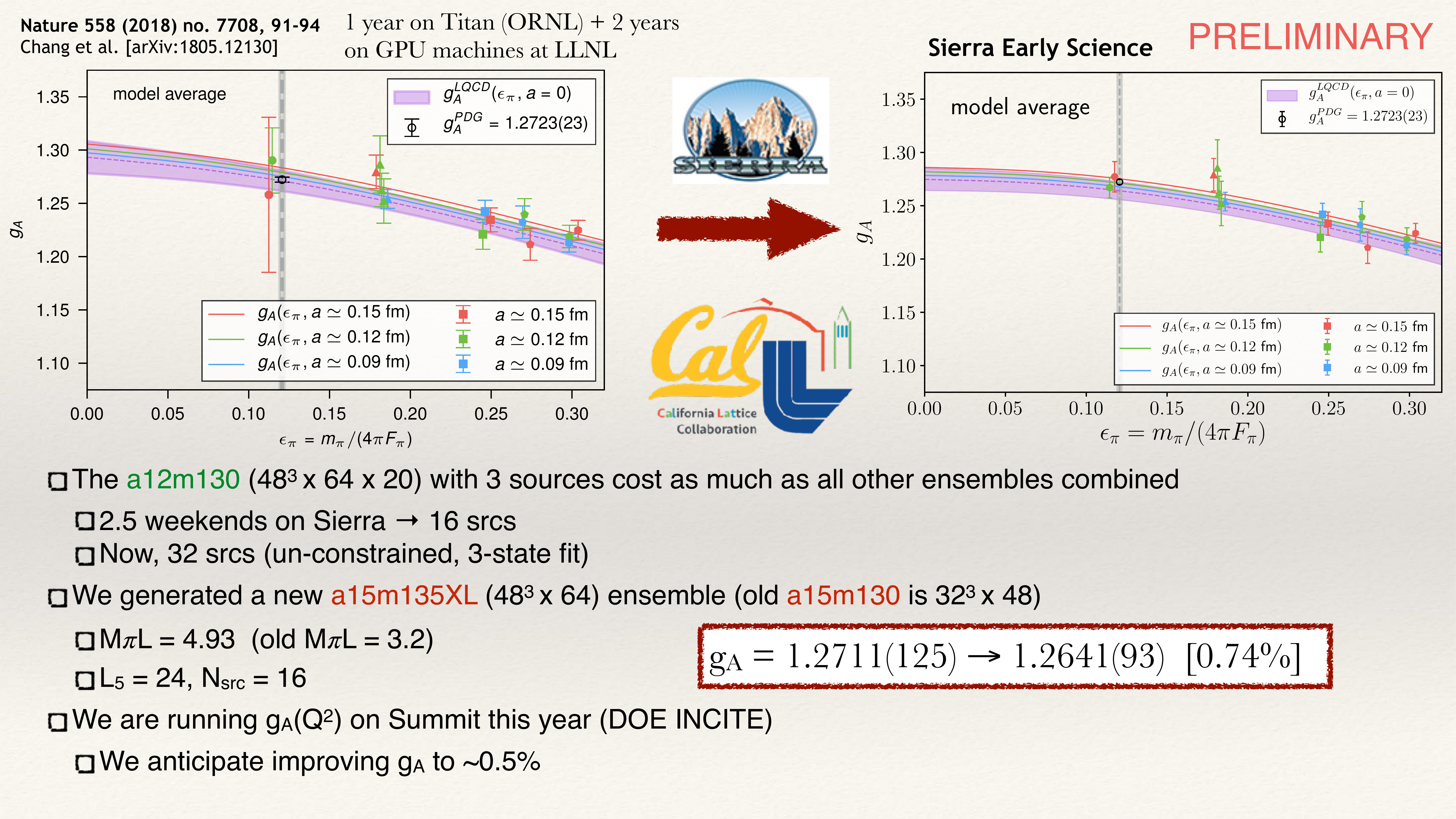}
\end{center}
\caption{A slide from the CalLat~\protect\cite{WalkerLoud.this} presentation explaining the improvements and expected reach of their calculations.}
\label{fig:CalLat}
\end{figure}
\begin{figure}
\begin{center}
\leavevmode
 \includegraphics[width=0.49\textwidth]{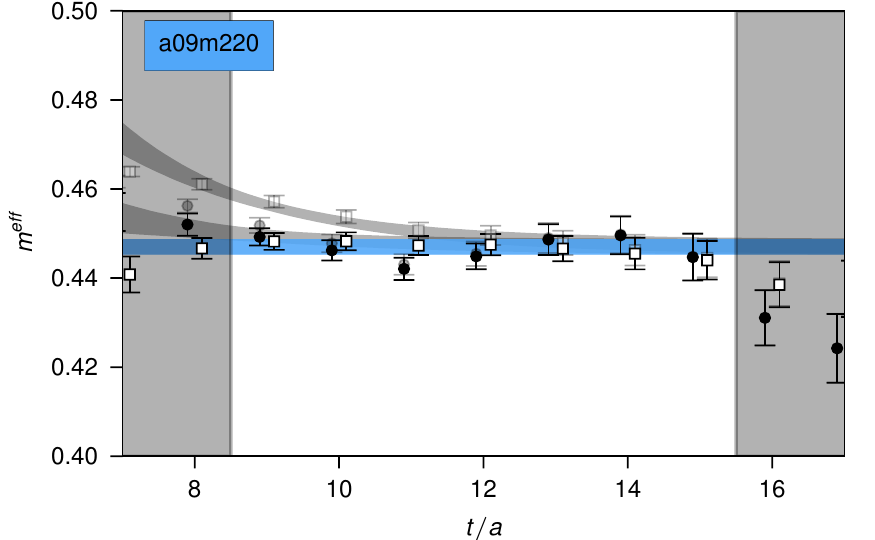}
 \includegraphics[width=0.49\textwidth]{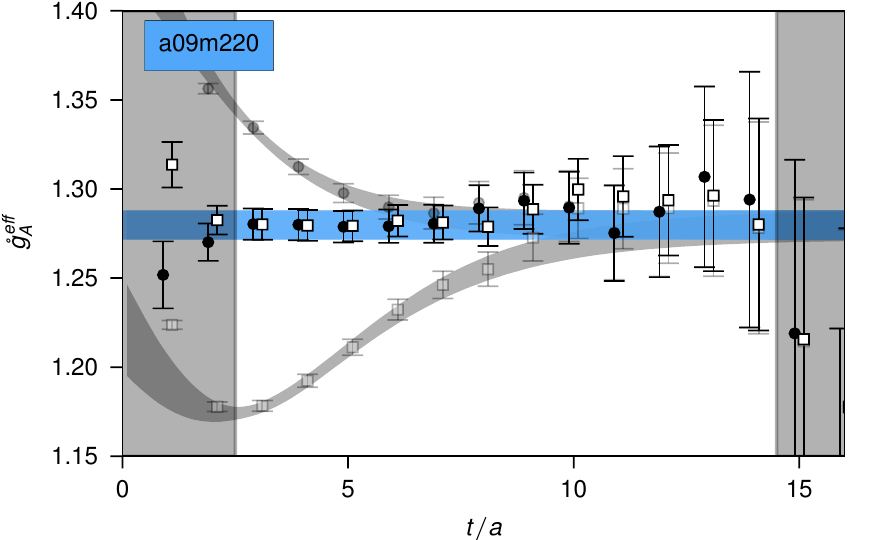}
\end{center}
\caption{Effective mass plots for the extraction of the ground-state mass (left) and axial charge \texorpdfstring{\(g_A\)}{g\string_A} (right) by the CalLat collaboration~\protect\cite{WalkerLoud.this}. The extraction of the charge is dominated by the low error points at \texorpdfstring{\(t/a\approx 3\)}{t/a\string~3}, the validity of which depends on reliable subtraction of excited states that clearly visible till \texorpdfstring{\(t/a\approx 10\)}{t/a\string~10}.}
\label{fig:CalLatmeff}
\end{figure}
\begin{figure}
\begin{tabular}{cc}
\includegraphics[width=0.45\textwidth]{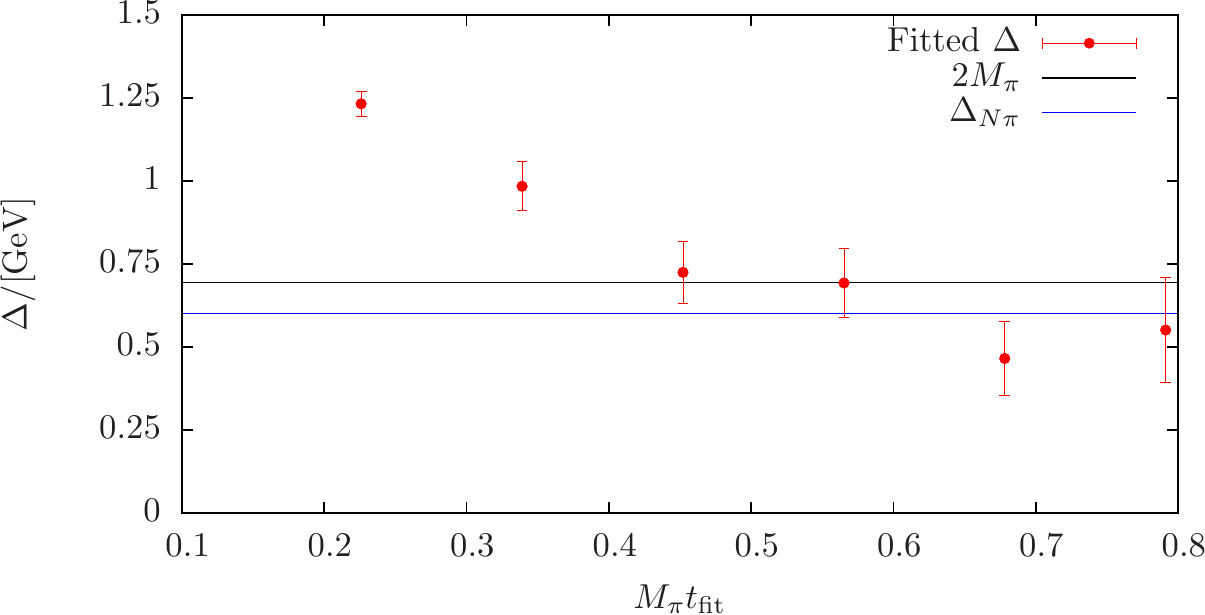}&
%
\includegraphics[width=0.4\textwidth]{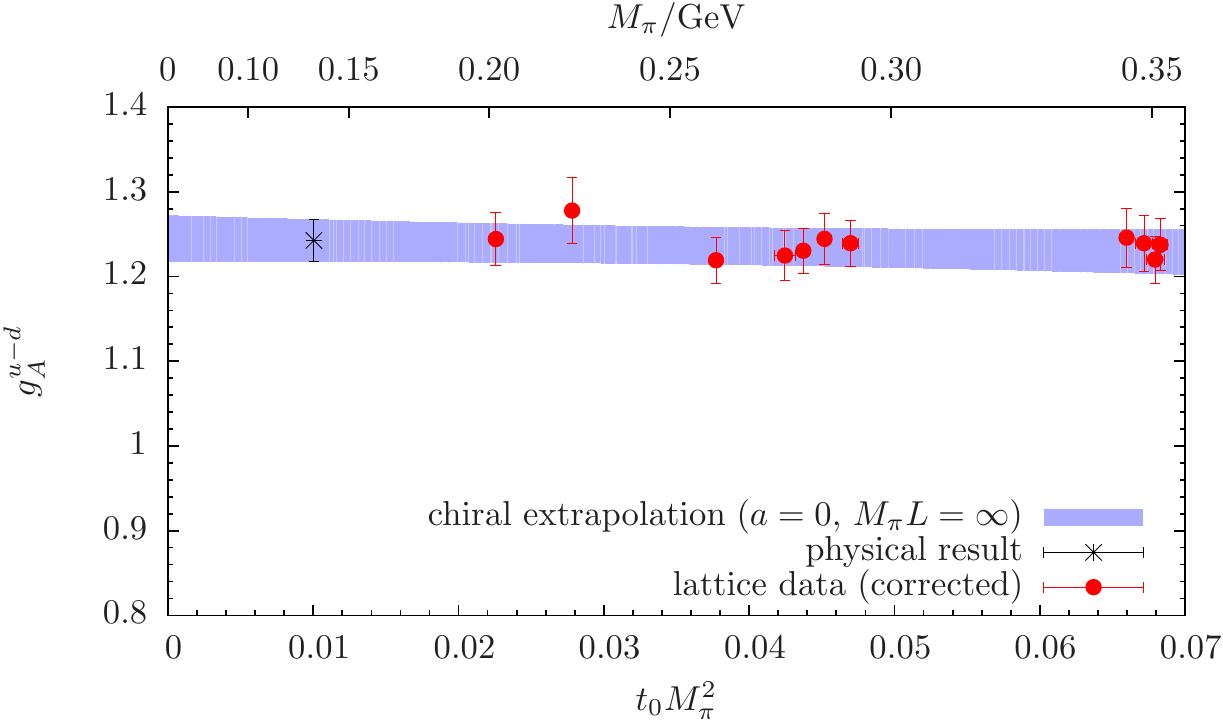}
\end{tabular}
\caption{Calculation of the axial charge \texorpdfstring{\(g_A\)}{g\string_A} by the Mainz collaboration~\protect\cite{Harris:2019bih}.  The left panel shows an example of the determination of excited-state mass gap and the right panel shows the chiral extrapolation of their data.  Chiral perturbation theory predicts the smallest mass gaps to be given by the horizontal lines, but the fitted mass gap is consistent with this only when they start at distances \texorpdfstring{\(M_\pi t_0 \approx 0.4\)}{M\string_pi t\string_0 \string~ 0.4}.}
\label{fig:Mainz}
\end{figure}

Fig.~\ref{fig:gAFLAG} from FLAG 2019~\cite{Aoki:2019cca} gives the
status of $g_A^{u-d}$ and summarizes that calculations have obtained
control at about the 3\% level.  Recent results from
LHPC~\cite{Hasan:2019noy} using 2+1-flavor Clover fermions,
RBC-UKQCD~\cite{Ohta.this} using Domain Wall fermions at a lattice
spacing of 1.73~GeV, and by PNDME collaboration using Clover-on-Clover
formulation with six 2+1 flavor clover fermions are consistent with
this picture. In contrast, as shown in Fig.~\ref{fig:CalLat}, the
CalLat collaboration is claiming sub-percent
accuracy~\cite{WalkerLoud.this}. The open question here is whether all
the systematic effects, in particular excited states, are accounted
for in their error estimate.  The reservation is illustrated in
Fig.~\ref{fig:CalLatmeff}. CalLat fit for $g_A$ on
the \(a\approx0.09\)~fm, \(M_\pi\approx220\)~MeV starts at \(t\sim 3a\)
and is dominated by the range $t/a=3$--$8$, however, the two point
function ($M^{\rm eff}$) shows clear indication of excited states even
up to \(t\sim 10a\).  In this regard, it is interesting to note the
observation of the Mainz group~\cite{Harris:2019bih} (See
Fig.~\ref{fig:Mainz}) that one needs to go out to about \(1/2M_\pi\)
before the lowest excited state mass expected from Chiral Perturbation
theory~\cite{Baer.this} is manifest.  If such low-mass states
contribute to \(g_A\), would it change the CalLat analysis based on
small values of $t$, in particular the error estimate, {\it i.e.\hbox{},} has the
statistical precision been traded for unaccounted-for systematics? Is
the CalLat analysis sensitive to possible low-lying excitations whose
contribution is only expected to manifest at large time separations?
Is the transition matrix element small?  More work is needed to
resolve these issues.

The ETMC collaboration~\cite{Alexandrou:2019brg} has compared various
strategies for estimating the excited state contamination: the plateau
method at large separations, the summation method, and using two-state
fits. They quote results of the two-state fit, which covers the
variation in the plateau values. Their results at the physical quark
mass, but at a single value of \(a\approx0.08\)~fm are consistent with
the CalLat determination, but with about 1.5\% errors. The ETMC result
does not, however, include the \(a\to0\) extrapolation
systematics. Lastly, the PACS collaboration has tested the efficacy of
a Coulomb-gauge exponential source versus the usual Gaussian-smeared
sources in removing excited state
contamination~\cite{Tsukamoto.this}. While the differences do not seem
to be major, their latest statement is that the Gaussian-smeared
sources are better.

\section{Isovector \texorpdfstring{\(g_S\) and \(g_T\)}{g\string_S and g\string_T}}
\begin{figure}
\begin{minipage}{0.6\linewidth}
  \tiny
  \setlength{\tabcolsep}{0pt}
\begin{tabular*}{\textwidth}[l]{l @{\extracolsep{\fill}} r lllllll l }
Collaboration & $\Nf$ & 
\hspace{1em}\begin{rotate}{90}{pub.}\end{rotate}\hspace{-1em} &
\hspace{1em}\begin{rotate}{90}{cont.}\end{rotate}\hspace{-1em} &
\hspace{1em}\begin{rotate}{90}{chiral}\end{rotate}\hspace{-1em}&
\hspace{1em}\begin{rotate}{90}{vol.}\end{rotate}\hspace{-1em}&
\hspace{1em}\begin{rotate}{90}{ren.}\end{rotate}\hspace{-1em}  &
\hspace{1em}\begin{rotate}{90}{states}\end{rotate}\hspace{-1em}  &
$g^{u-d}_S$\\
\hline
\hline
%
PNDME 18 &  2+1+1 & \gA & \good$^\ddag$ & \good & \good & \good & \good & 1.022(80)(60) \\[0ex]
PNDME 16 &  2+1+1 & \gA & \soso$^\ddag$ & \good & \good & \good & \good & 0.97(12)(6) \\[0ex]
PNDME 13 &  2+1+1 & \gA & \bad$^\ddag$ & \bad & \good & \good & \good & 0.72(32) \\[0ex]
\hline
Mainz 18 &  2+1 & \rC & \good & \soso & \good & \good & \good & 1.22(11) \\[0ex]
JLQCD 18 &  2+1 & \gA & \bad & \soso & \soso & \good & \good & 0.88(8)(3)(7) \\[0ex]
LHPC 12 &  2+1 & \gA & \bad$^\ddag$ & \good & \good & \good & \good & 1.08(28)(16) \\[0ex]
\hline
ETM 17 &  2 & \gA & \bad & \soso & \soso & \good & \good & 0.930(252)(48)(204) \\[0ex]
RQCD 14 &  2 & \gA & \soso & \good & \good & \good & \bad & 1.02(18)(30) \\[0ex]
\hline
\hline
\end{tabular*}
\tiny
{
$^\ddag$Not fully O(a) improved.
}
\end{minipage}%
\begin{minipage}{0.4\linewidth}
  \hbox to 0pt{\includegraphics[width=\linewidth, height=7\baselineskip]{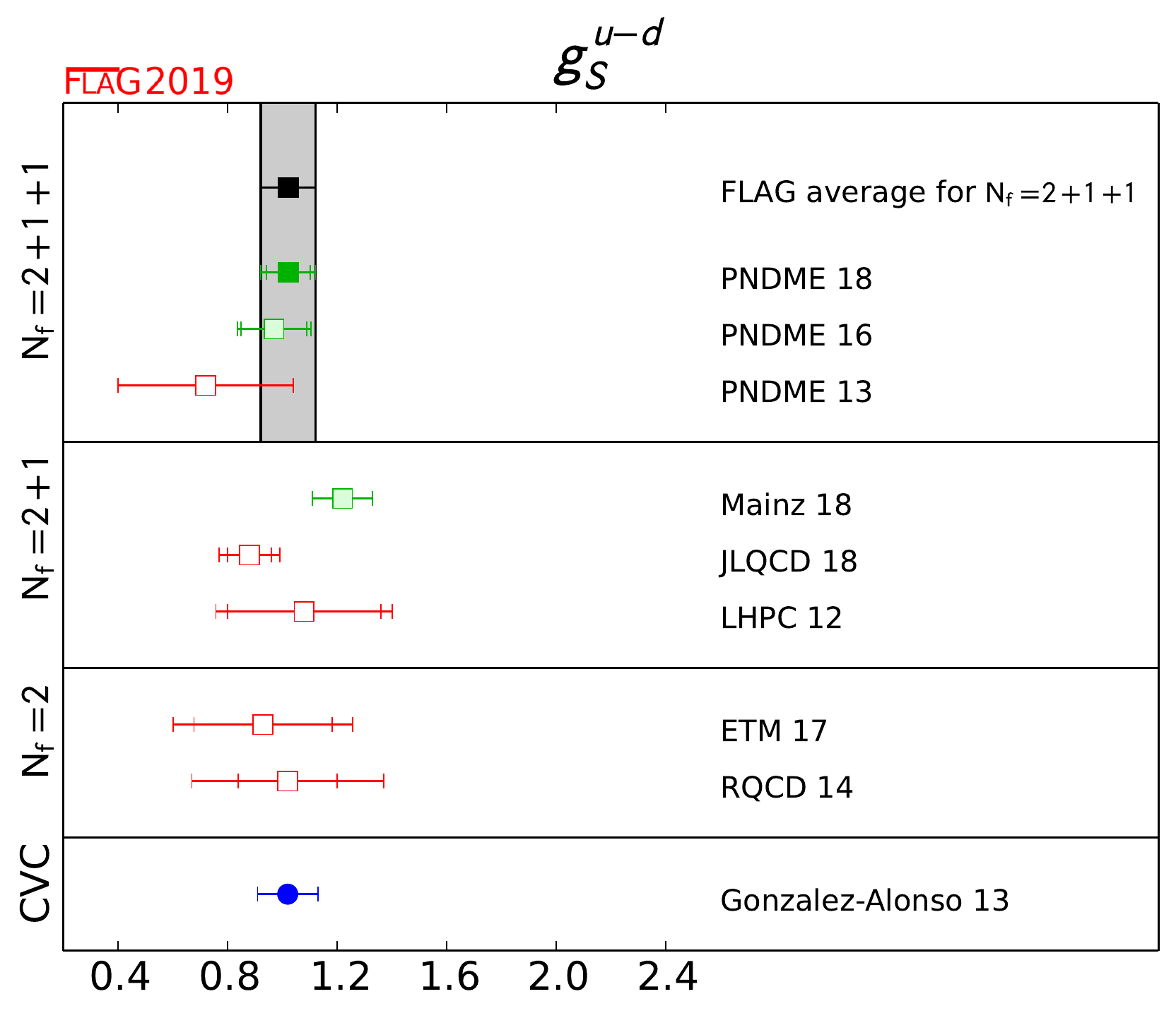}\hss}
\end{minipage}

  \caption{Compilation of the world data on isovector \texorpdfstring{\(g_S\)}{g\string_S} presented in the FLAG~\protect\cite{Aoki:2019cca} Review.}
  \label{fig:gSFLAG}
\end{figure}

\begin{figure}
\begin{minipage}{0.6\linewidth}
\tiny
  \setlength{\tabcolsep}{0pt}
\begin{tabular*}{\textwidth}[l]{l @{\extracolsep{\fill}} r lllllll l }
Collaboration & $\Nf$ & 
\hspace{1em}\begin{rotate}{90}{pub.}\end{rotate}\hspace{-1em} &
\hspace{1em}\begin{rotate}{90}{cont.}\end{rotate}\hspace{-1em} &
\hspace{1em}\begin{rotate}{90}{chiral}\end{rotate}\hspace{-1em}&
\hspace{1em}\begin{rotate}{90}{vol.}\end{rotate}\hspace{-1em}&
\hspace{1em}\begin{rotate}{90}{ren.}\end{rotate}\hspace{-1em}  &
\hspace{1em}\begin{rotate}{90}{states}\end{rotate}\hspace{-1em}  &
$g^{u-d}_T$\\
\hline
\hline
%
PNDME 18 & \c 2+1+1 & \gA & \good$^\ddag$ & \good & \good & \good & \good & 0.989(32)(10) \\[0ex]
PNDME 16 & \c 2+1+1 & \gA & \soso$^\ddag$ & \good & \good & \good & \good & 0.987(51)(20) \\[0ex]
PNDME 15 & \c 2+1+1 & \gA & \soso$^\ddag$ & \good & \good & \good & \good & 1.020(76) \\[0ex]
PNDME 13 & \c 2+1+1 & \gA & \bad$^\ddag$ & \bad & \good & \good & \good & 1.047(61) \\[0ex]
\hline
Mainz 18 & \c 2+1 & \rC & \good & \soso & \good & \good & \good & 0.979(60) \\[0ex]
JLQCD 18 & \c 2+1 & \gA & \bad & \soso & \soso & \good & \good & 1.08(3)(3)(9) \\[0ex]
LHPC 12 & \c 2+1 & \gA & \bad$^\ddag$ & \good & \good & \good & \good & 1.038(11)(12) \\[0ex]
RBC/UKQCD 10D & \c 2+1 & \gA & \bad & \bad & \soso & \good & \bad & 0.9(2) \\[0ex]
\hline
ETM 17 & \c 2 & \gA & \bad & \soso & \soso & \good & \good & 1.004(21)(2)(19) \\[0ex]
ETM 15D & \c 2 & \gA & \bad & \soso & \soso & \good & \good & 1.027(62) \\[0ex]
RQCD 14 & \c 2 & \gA & \soso & \good & \good & \good & \bad & 1.005(17)(29) \\[0ex]
RBC 08 & \c 2 & \gA & \bad & \bad & \bad & \good & \bad & 0.93(6) \\[0ex]
\hline
\hline
\end{tabular*}
\tiny
{
$^\ddag$Not fully O(a) improved.
}
\end{minipage}
\begin{minipage}{0.4\linewidth}
  \vspace*{\baselineskip}
  \hbox to 0pt{\includegraphics[width=0.99\linewidth,height=12\baselineskip]{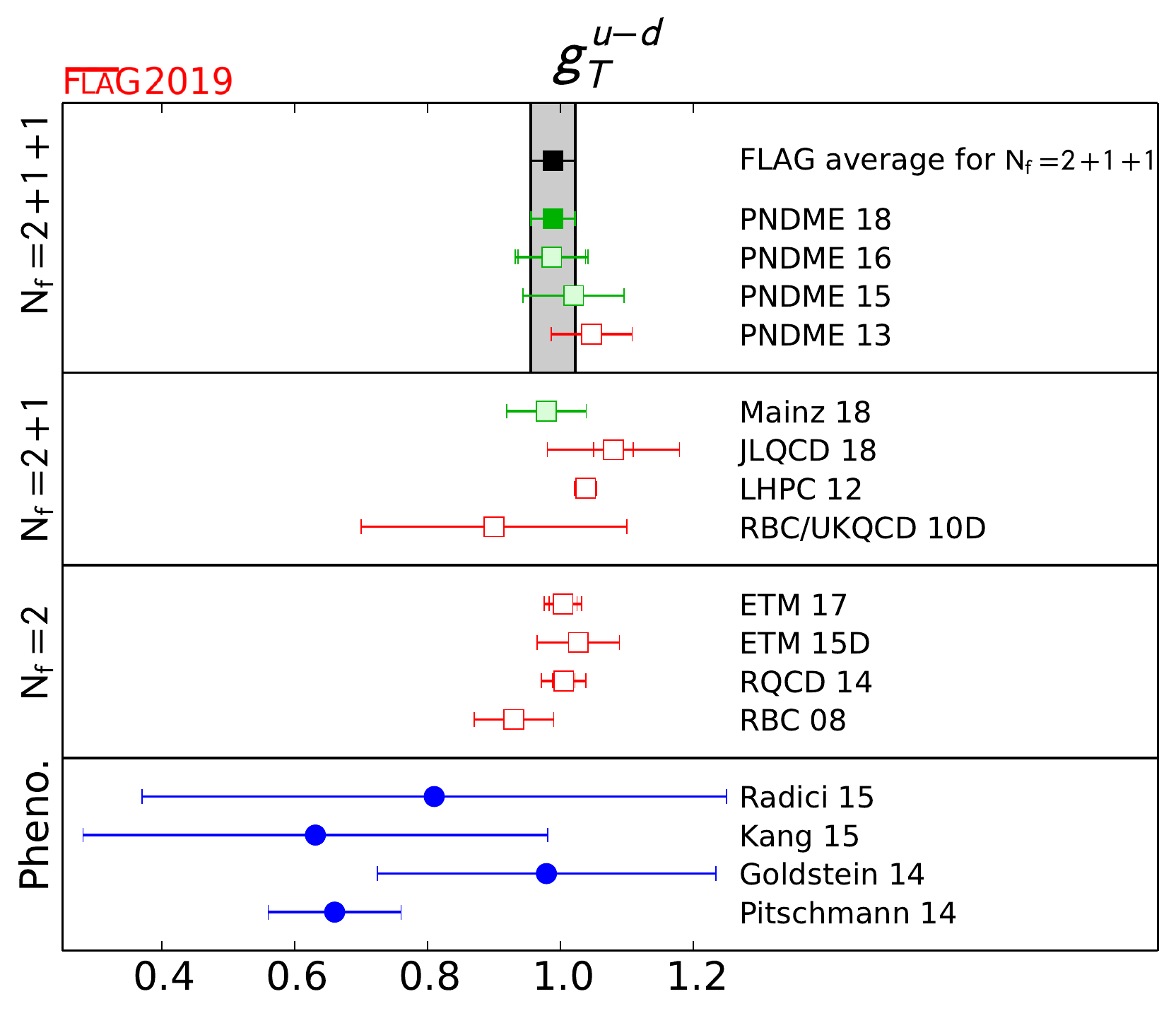}\hss}
\end{minipage}

  \caption{Compilation of the world data on isovector \texorpdfstring{\(g_T\)}{g\string_T} presented in the FLAG~\protect\cite{Aoki:2019cca} Review.}
  \label{fig:gTFLAG}
\end{figure}

According to FLAG 2019~\cite{Aoki:2019cca}, the isovector
\(g_S=1.022(80)(60)\) is currently known to about 10\% and
\(g_T=0.989(32)(10)\) to about 4\%, see Figs.~\ref{fig:gSFLAG} and
\ref{fig:gTFLAG}. New calculations presented in this conference from
the LHPC~\cite{Hasan:2019noy}, PACS~\cite{Tsukamoto.this},
Mainz~\cite{Harris:2019bih} and PNDME~\cite{Park.this} collaborations
agree with this consensus within one standard deviation.  The ETMC
collaboration~\cite{Alexandrou:2019brg} again finds that the two-state
fits to remove excited state contributions resulted in the smallest
statistical errors. Their results, \(1.35(17)\), at a single lattice spacing of
\(a\approx 0.08~\rm fm\) differ from the world averages of the
\(a\to0\) extrapolated values by about two standard deviations. Note, however, the
systematic due to $a \to 0$ extrapolation in their estimate from this one
ensemble is an unknown.

\section{Flavor Diagonal Charges}

\begin{figure}
\begin{minipage}{\linewidth}
\tiny
  \setlength{\tabcolsep}{0pt}
\begin{tabular*}{\textwidth}[l]{l @{\extracolsep{\fill}} r lllllll l }
Collaboration & $\Nf$ & 
\hspace{1em}\begin{rotate}{90}{pub.}\end{rotate}\hspace{-1em} &
\hspace{1em}\begin{rotate}{90}{cont.}\end{rotate}\hspace{-1em} &
\hspace{1em}\begin{rotate}{90}{chiral}\end{rotate}\hspace{-1em}&
\hspace{1em}\begin{rotate}{90}{vol.}\end{rotate}\hspace{-1em}&
\hspace{1em}\begin{rotate}{90}{ren.}\end{rotate}\hspace{-1em}  &
\hspace{1em}\begin{rotate}{90}{states}\end{rotate}\hspace{-1em}  &
$\Delta u$ & $\Delta d$ \\
\hline
\hline
%
PNDME 18A &  2+1+1 & \gA & \good$^\ddag$ & \good & \good & \good & \good & 0.777(25)(30)$^\#$ & $-$0.438(18)(30)$^\#$ \\[0.5ex]
\hline
$\chi$QCD 18 &  2+1 & \gA & \soso & \good & \good & \good & \good & 0.847(18)(32)$^\$$ & $-$0.407(16)(18)$^\$$ \\[0.5ex]
\hline
ETM 17C &  2 & \gA & \bad & \soso & \soso & \good & \good & $0.830(26)(4)$ & $-0.386(16)(6)$ \\[0.5ex]
\hline
\hline
 & & & & & & & & $\Delta s$& \\[-0cm]
\hline
\hline
PNDME 18A &  2+1+1 & \gA & \good$^\ddag$ & \good & \good & \good & \good &  $-$0.053(8)$^\#$ & \\[0.5ex]
\hline
$\chi$QCD 18 &  2+1 & \gA & \soso & \good & \good & \good & \good &  $-$0.035(6)(7)$^\$$ & \\[0.5ex]
JLQCD 18 &  2+1 & \gA & \bad & \soso & \soso & \good & \good &  $-$0.046(26)(9)$^{\#}$ & \\[0.5ex]
$\chi$QCD 15 &  2+1 & \gA & \bad & \soso & \bad & \good & \good &  $-$0.0403(44)(78)$^\#$ & \\[0.5ex]
Engelhardt 12 &  2+1 & \gA & \bad & \soso & \bad & \good & \good &  $-$0.031(17)$^\#$ & \\[0.5ex]
\hline
ETM 17C &  2 & \gA & \bad & \soso & \soso & \good & \good &  $-$0.042(10)(2) & \\[0.5ex]
\hline
\hline
\end{tabular*}
\tiny
{
$^\#$$Z_A^{n.s.}=Z_A^{s}$ assumed.
$^\$$Also partially quenched analysis.
$^\ddag$Not fully O(a) improved.
}
\end{minipage}%
\vspace*{0pt}
\begin{minipage}{\linewidth}
  \hbox to 0pt{\includegraphics[width=0.33\linewidth]{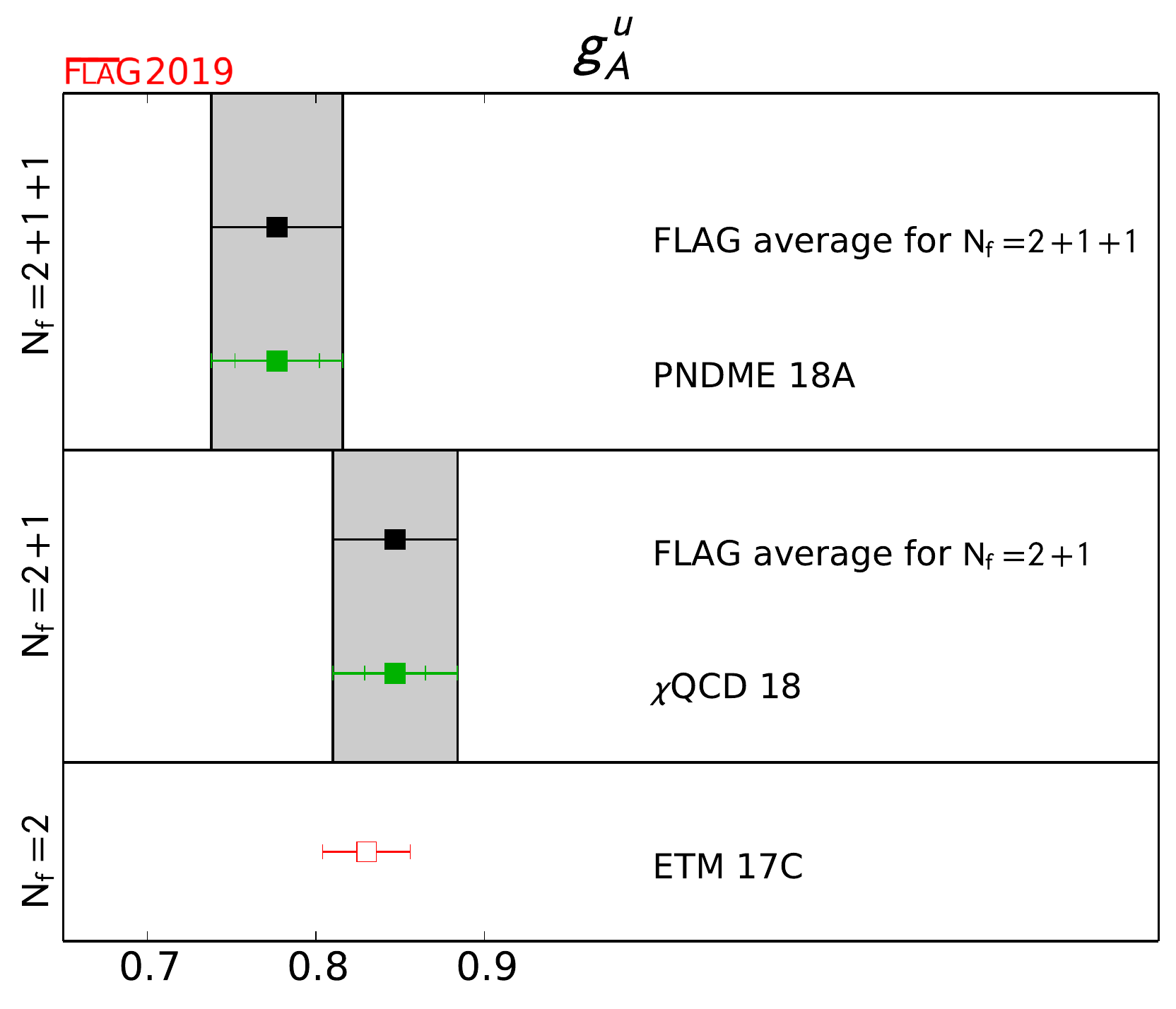}
               \includegraphics[width=0.33\linewidth]{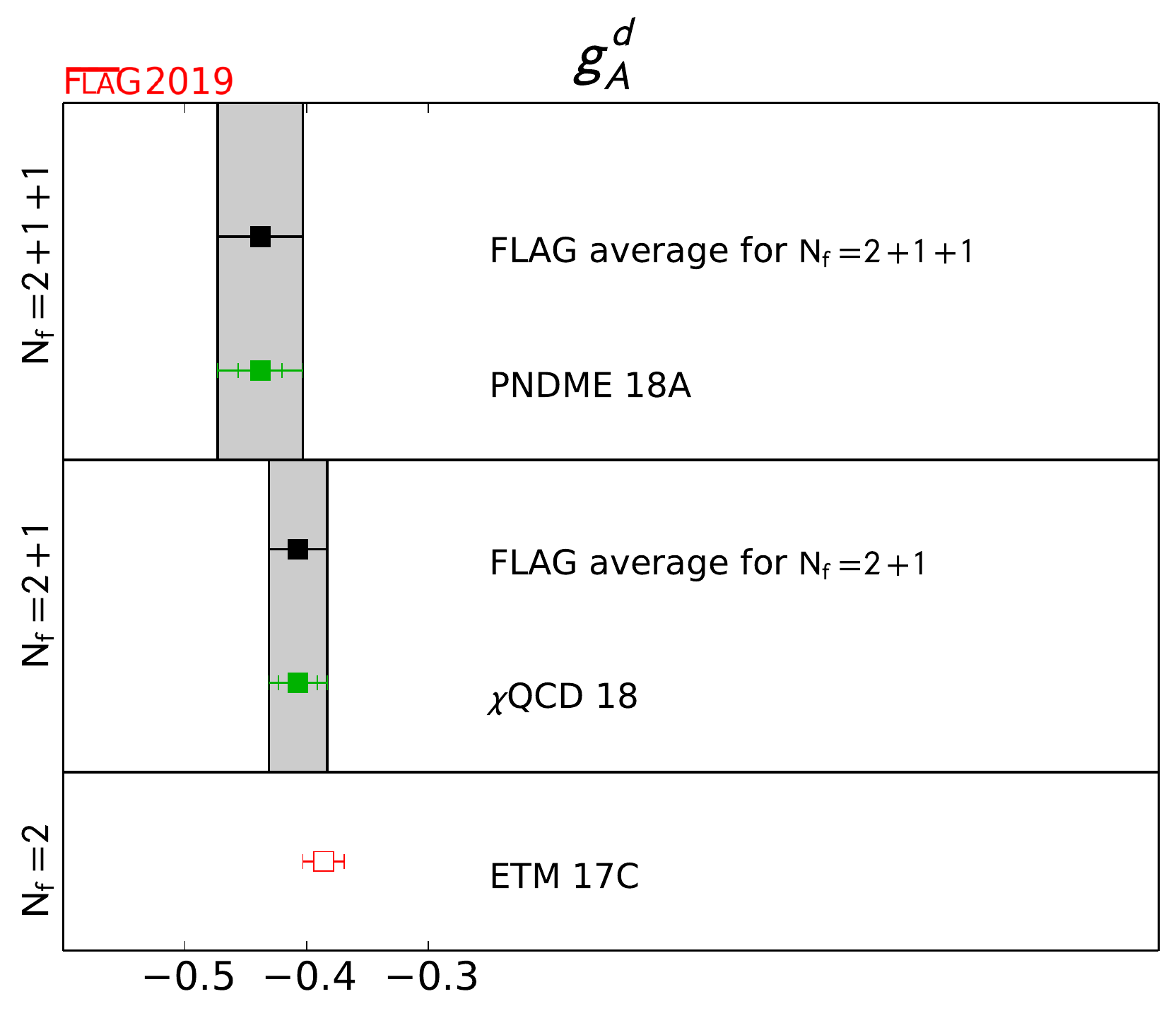}
               \includegraphics[width=0.33\linewidth]{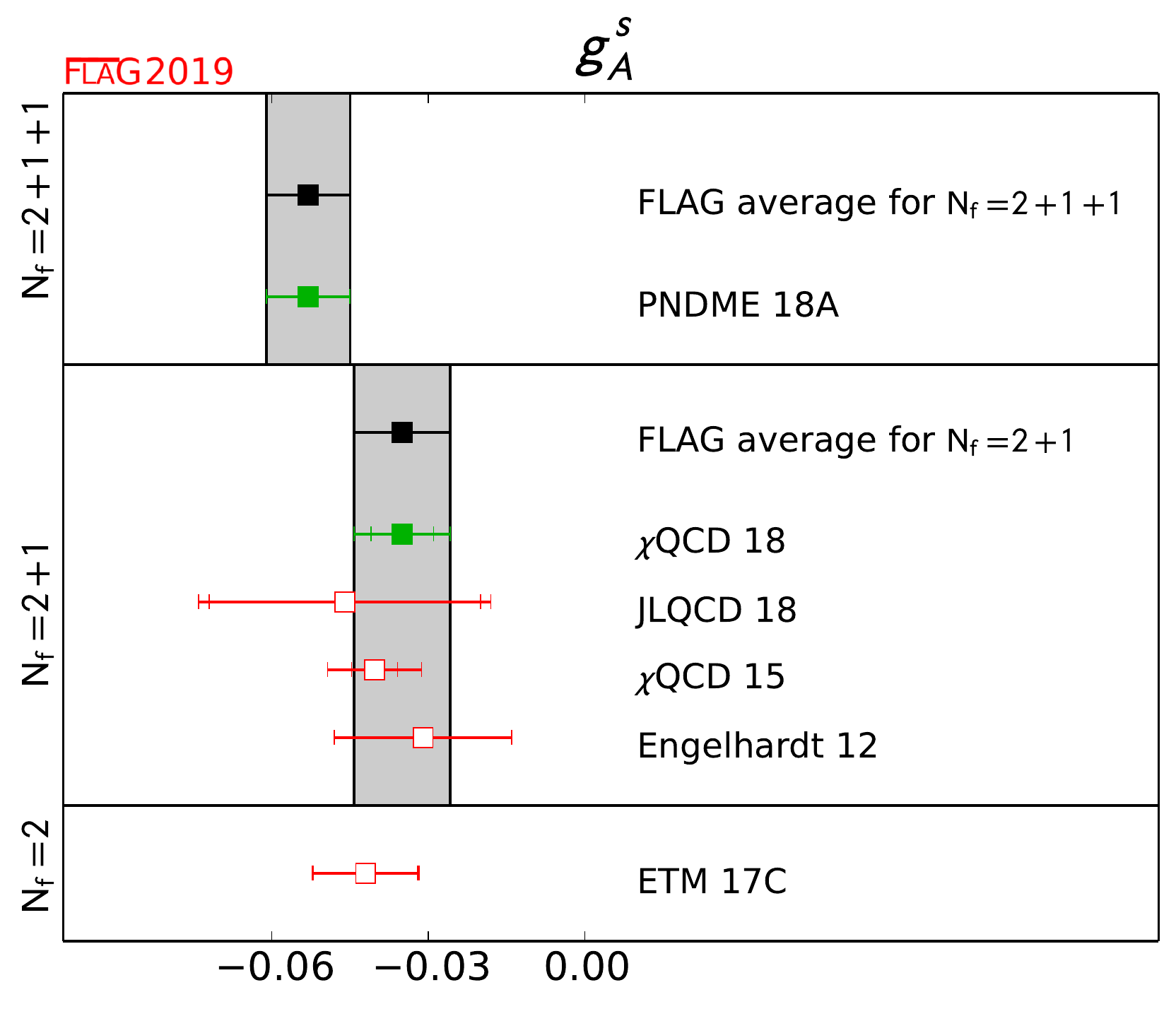}\hss}
\end{minipage}

  \caption{Compilation of the world data on flavor-diagonal \texorpdfstring{\(g_A\)}{g\string_A} presented in the FLAG~\protect\cite{Aoki:2019cca} Review.}
  \label{fig:gAdiagFLAG}
\end{figure}

Estimates for the flavor-diagonal charges have larger statistical and
systematic uncertainty than the isovector case. Also, the connected and
disconnected contributions are fit separately (implying a partially
quenched analysis) to remove the excited-state
contamination~\cite{Lin:2018obj}. The associated, presumably small,
systematic is not estimated. As shown in Fig.~\ref{fig:gAdiagFLAG}, the \(u\) and \(d\)
quark contributions, $g_A^u$ and $g_A^d$ are determined at the 5 and
8\% accuracy. The difference is due to the fact that the magnitude of
the \(d\) quark contribution is smaller but has a similar error.  The
much smaller strange quark contribution is known at 15\% level. So
far, very few collaborations have attempted to control all the
systematics in the calculation of flavor diagonal charges, and a
dependence on the sea charm quark cannot yet be ruled out. New results
from the Mainz~\cite{Harris:2019bih} and PNDME~\cite{Park.this}
collaboration agree with the previous 2+1 and 2+1+1 results (FLAG
averages), respectively.

\begin{figure}
\begin{minipage}{\linewidth}
\tiny
  \setlength{\tabcolsep}{0pt}
\begin{tabular*}{\textwidth}[l]{l @{\extracolsep{\fill}} r lllllll l }
Collaboration & $\Nf$ & 
\hspace{1em}\begin{rotate}{90}{pub.}\end{rotate}\hspace{-1em} &
\hspace{1em}\begin{rotate}{90}{cont.}\end{rotate}\hspace{-1em} &
\hspace{1em}\begin{rotate}{90}{chiral}\end{rotate}\hspace{-1em}&
\hspace{1em}\begin{rotate}{90}{vol.}\end{rotate}\hspace{-1em}&
\hspace{1em}\begin{rotate}{90}{ren.}\end{rotate}\hspace{-1em}  &
\hspace{1em}\begin{rotate}{90}{states}\end{rotate}\hspace{-1em}  &
$g_T^u$ & $g_T^d$ \\
\hline
\hline
%
PNDME 18B &  2+1+1 & \oP & \good$^\ddag$ & \good & \good & \good & \good & 0.784(28)(10)$^\#$ & $-$0.204(11)(10)$^\#$ \\[0ex]
PNDME 16 &  2+1+1 & \gA & \soso$^\ddag$ & \good & \good & \good & \good & 0.792(42)$^{\#\&}$ & $-$0.194(14)$^{\#\&}$ \\[0ex]
PNDME 15 &  2+1+1 & \gA & \soso$^\ddag$ & \good & \good & \good & \good & 0.774(66)$^\#$ & $-$0.233(28)$^\#$ \\[0ex]
\hline
JLQCD 18 &  2+1 & \gA & \bad & \soso & \soso & \good & \good & 0.85(3)(2)(7) & $-$0.24(2)(0)(2) \\[0ex]
\hline
ETM 17 &  2 & \gA & \bad & \soso & \soso & \good & \good & 0.782(16)(2)(13) & $-$0.219(10)(2)(13) \\[0ex]
\hline
\hline
 & & & & & & & & $ g_T^s $& \\
\hline
\hline
PNDME 18B &  2+1+1 & \oP & \good$^\ddag$ & \good & \good & \good & \good &  $-$0.0027(16)$^\#$ & \\[0ex]
PNDME 15 &  2+1+1 & \gA & \soso$^\ddag$ & \good & \good & \good & \good &  0.008(9)$^\#$ & \\[0ex]
\hline
JLQCD 18 &  2+1 & \gA & \bad & \soso & \soso & \good & \good &  $-$0.012(16)(8) & \\[0ex]
\hline
ETM 17 &  2 & \gA & \bad & \soso & \soso & \good & \good &  $-$0.00319(69)(2)(22) & \\[0ex]
\hline
\hline
\end{tabular*}
\tiny
{
$^\#$$Z_T^{n.s.}=Z_T^{s}$ assumed. 
$^\&$Only `connected'.
$^\ddag$Not fully O(a) improved.
}
\end{minipage}
\vspace*{0pt}
\begin{minipage}{\linewidth}
  \hbox to 0pt{\includegraphics[width=0.33\linewidth]{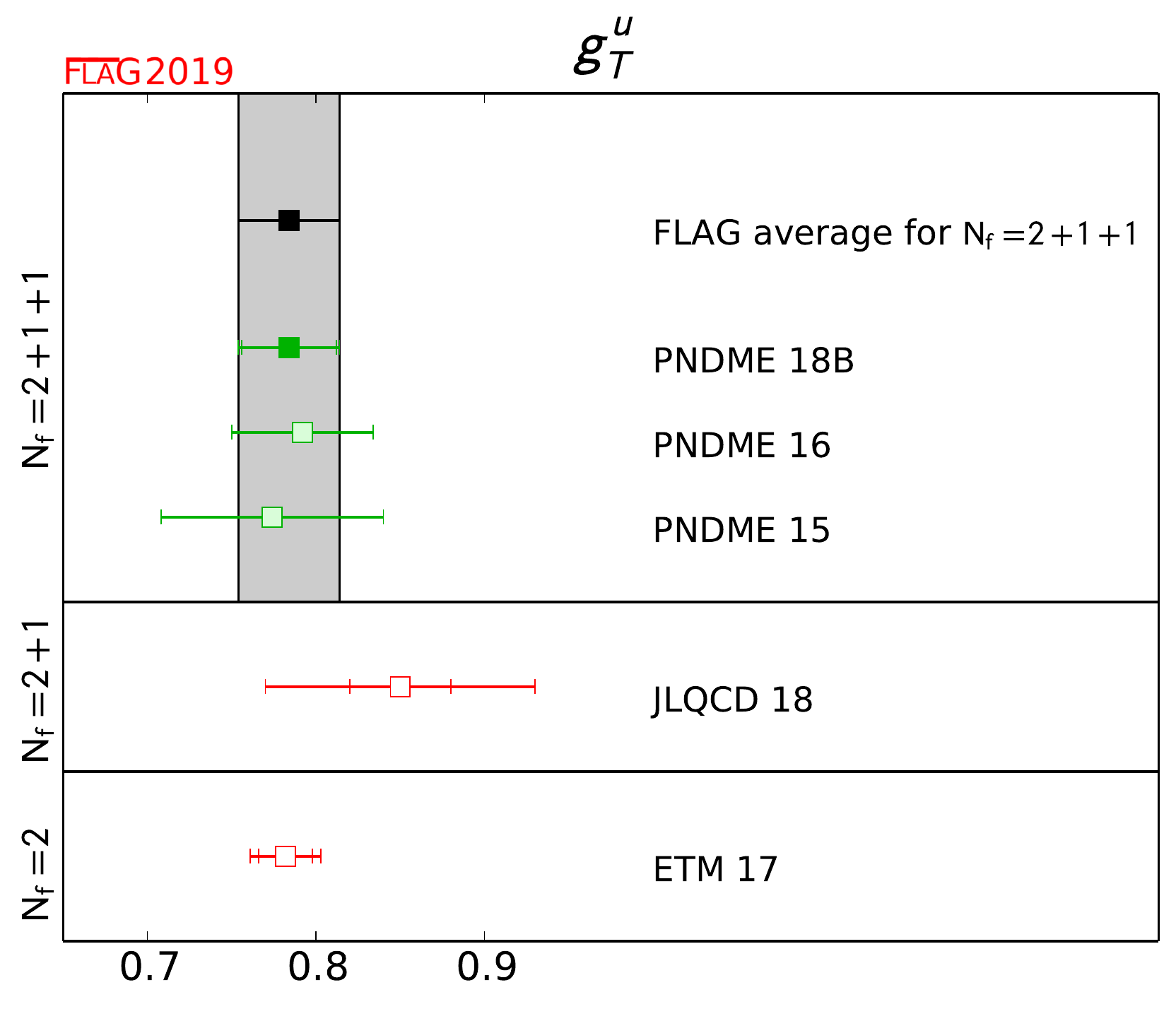}
               \includegraphics[width=0.33\linewidth]{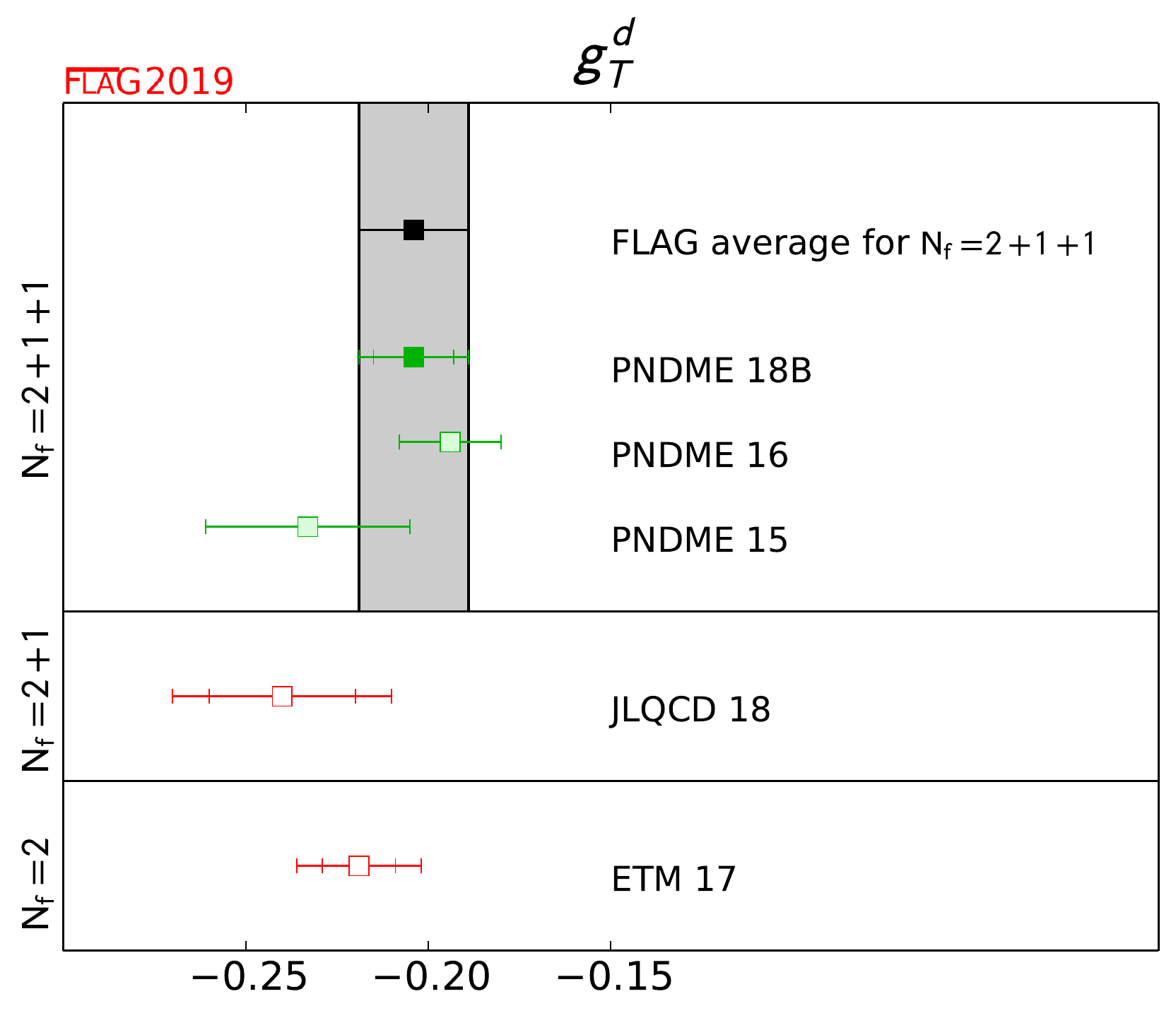}
               \includegraphics[width=0.33\linewidth]{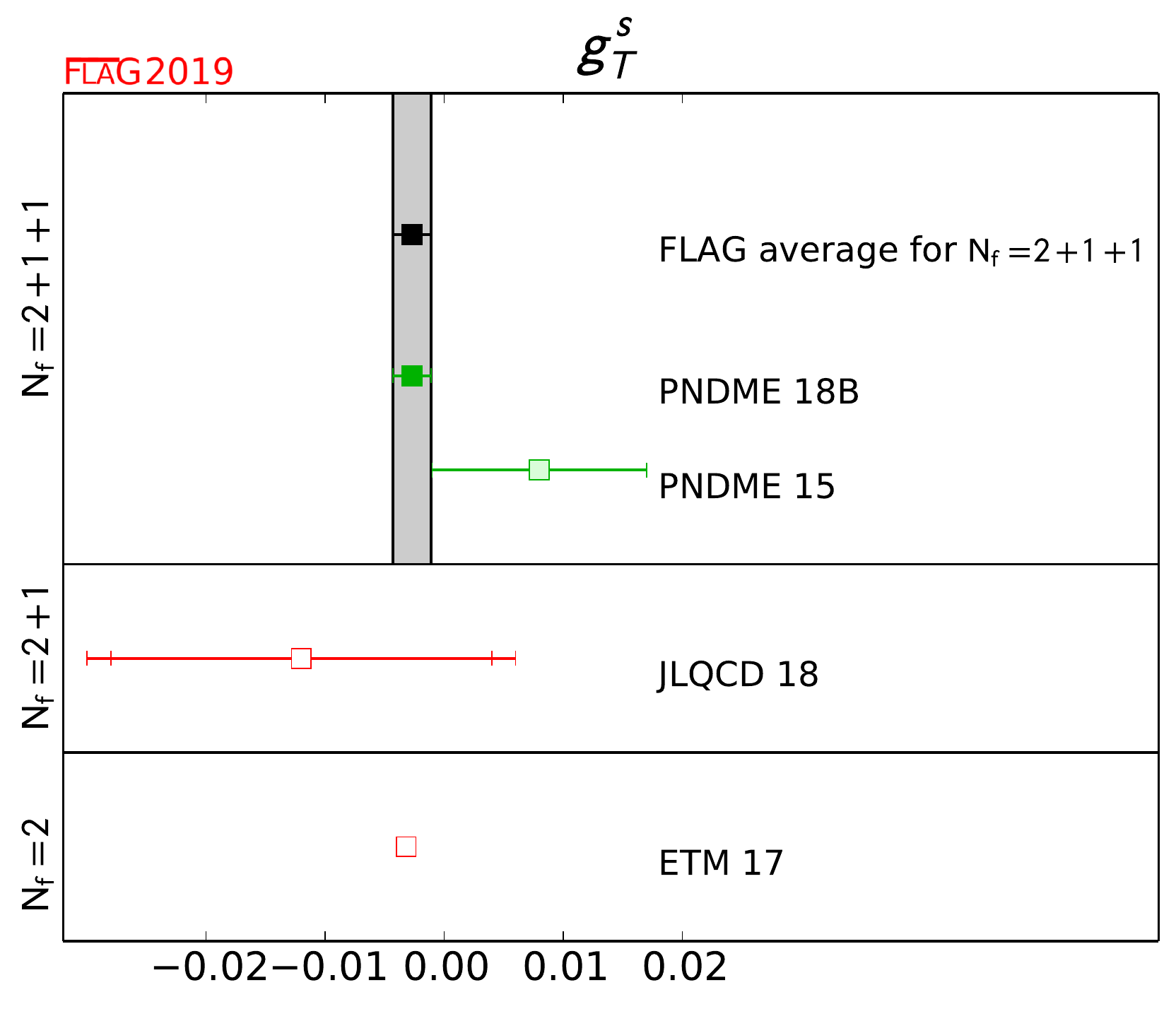}\hss}
\end{minipage}

  \caption{Compilation of the world data on flavor-diagonal \texorpdfstring{\(g_T\)}{g\string_T} presented in the FLAG~\protect\cite{Aoki:2019cca} Review.}
  \label{fig:gTdiagFLAG}
\end{figure}

The situation of the flavor-diagonal tensor charges is similar.  As
per the FLAG averages (Fig.~\ref{fig:gTdiagFLAG}), the \(u\) and \(d\)
quark contributions are known at 4 and 7\% accuracy, whereas the \(s\)
quark contribution differs from zero only by \(1.7\sigma\). New
calculations from the Mainz~\cite{Djukanovic.this} and
PNDME~\cite{Park.this} collaborations give consistent values for the
\(u\) and \(d\) quark contributions. PNDME have reduced the
uncertainty on the \(s\) quark matrix element, $g_T^s$, and it is now negative
with ${}>2\sigma$ confidence.

\begin{figure}
\begin{minipage}{\linewidth}
\tiny
  \setlength{\tabcolsep}{0pt}
\begin{tabular*}{\textwidth}[l]{l @{\extracolsep{\fill}} r lllllll l }
Collaboration & $\Nf$ & 
\hspace{1em}\begin{rotate}{90}{pub.}\end{rotate}\hspace{-1em} &
\hspace{1em}\begin{rotate}{90}{cont.}\end{rotate}\hspace{-1em} &
\hspace{1em}\begin{rotate}{90}{chiral}\end{rotate}\hspace{-1em}&
\hspace{1em}\begin{rotate}{90}{vol.}\end{rotate}\hspace{-1em}&
\hspace{1em}\begin{rotate}{90}{ren.}\end{rotate}\hspace{-1em}  &
\hspace{1em}\begin{rotate}{90}{states}\end{rotate}\hspace{-1em}  &
$\sigma_{\pi N}$~[MeV] & $\sigma_s$~[MeV] \\
\hline
\hline
%
MILC 12C &  2+1+1 & \gA & \good & \good & \good & \good & $-$ & 0.44(8)(5)$\times m_s$$^{\P\S}$ \\[0ex]
\hline
JLQCD 18 &  2+1 & \gA & \bad & \soso & \soso & \good & 26(3)(5)(2) & 17(18)(9) \\[0ex]
$\chi$QCD 15A &  2+1 & \gA & \soso & \good & \good & \good & 45.9(7.4)(2.8)$^\$$ & 40.2(11.7)(3.5)$^\$$ \\[0ex]
$\chi$QCD 13A &  2+1 & \gA & \bad & \bad & \soso & \good & $-$ & 33.3(6.2)$^\$$ \\[0ex]
JLQCD 12A &  2+1 & \gA & \bad & \soso & \soso & \good & $-$ & 0.009(15)(16)$\times m_N$$^\dagger$ \\[0ex]
Engelhardt 12 &  2+1 & \gA & \bad & \soso & \bad & \good & $-$ & 0.046(11)$\times m_N$$^\dagger$ \\[0ex]
MILC 12C &  2+1 & \gA & \good & \soso & \good &  \good & $-$ &0.637(55)(74)$\times m_s$$^{\P\S}$ \\[0ex]
MILC 09D &  2+1 & \gA & \good & \soso & \good & \good & $-$ &59(6)(8)$^\S$ \\[0ex]
\hline
ETM 16A &  2 & \gA & \bad & \soso & \soso & \good & 37.2(2.6)($^{4.7}_{2.9}$) & 41.1(8.2)($^{7.8}_{5.8}$) \\[0ex]
RQCD 16 &  2 & \gA & \soso & \good & \good & \bad & 35(6) & 35(12) \\[0ex]
\hline
\hline
ETM 14A &  2+1+1 & \gA & \good & \soso & \soso && 64.9(1.5)(13.2)$^\triangle$ & $-$ \\[0ex]
\hline
BMW 15 &  2+1 & \gA & \good$^\ddag$ & \good & \good && 38(3)(3) & 105(41)(37) \\[0ex]
Junnarkar 13 &  2+1 & \gA & \soso & \soso & \soso && $-$ & 48(10)(15) \\[0ex]
Shanahan 12 &  2+1 & \gA & \bad & \soso & \soso && 45(6)/51(7)$^\star$ & 21(6)/59(6)$^\star$ \\[0ex]
JLQCD 12A &  2+1 & \gA & \bad & \soso & \soso && $-$ & 0.023(29)(28)$\times m_N$$^\dagger$ \\[0ex]
QCDSF 11 &  2+1 & \gA & \bad & \bad & \soso && 31(3)(4) & 71(34)(59) \\[0ex]
BMW 11A &  2+1 & \gA & \soso$^\ddag$ & \good & \soso && 39(4)($^{18}_{7}$) & 67(27)($^{55}_{47}$) \\[0ex]
Martin~Camalich 10 &  2+1 & \gA & \bad & \good & \bad && 59(2)(17) & $-$4(23)(25) \\[0ex]
PACS-CS 09 &  2+1 & \gA & \bad & \good & \bad && 75(15) & $-$ \\[0ex]
\hline
QCDSF 12 &  2 & \gA & \soso & \good & \soso && 37(8)(6) & $-$ \\[0ex]
JLQCD 08B &  2 & \gA & \bad & \soso & \bad && 53(2)($^{+21}_{-7}$) & $-$ \\[0ex]
\hline
\hline
\end{tabular*}
\tiny
$^\triangle$Multiple results.
$^\ddag$Not fully O(a) improved.
$^\star$Two results are quoted.
$^\dagger$From $f_{T_s}=\sigma_s/m_N$
$^\$$Also partially quenched
$^\S$Uses a hybrid method
$^\P$At $\mu=2$~GeV in $\msbar$ scheme.
\end{minipage}%
\vspace*{0pt}
\begin{minipage}{\linewidth}
\hbox to 0pt{\includegraphics[width=0.5\linewidth]{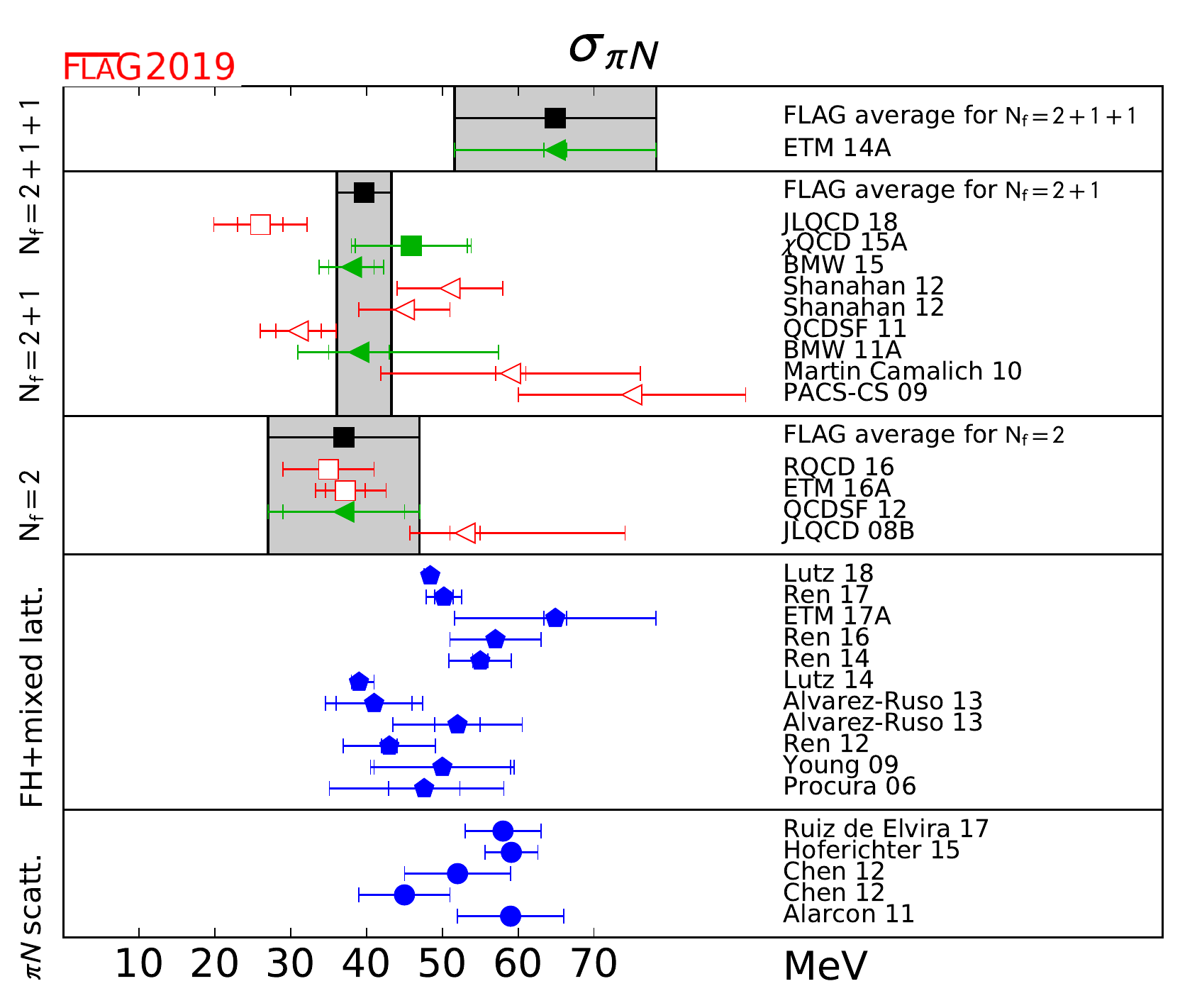}%
             \includegraphics[width=0.5\linewidth]{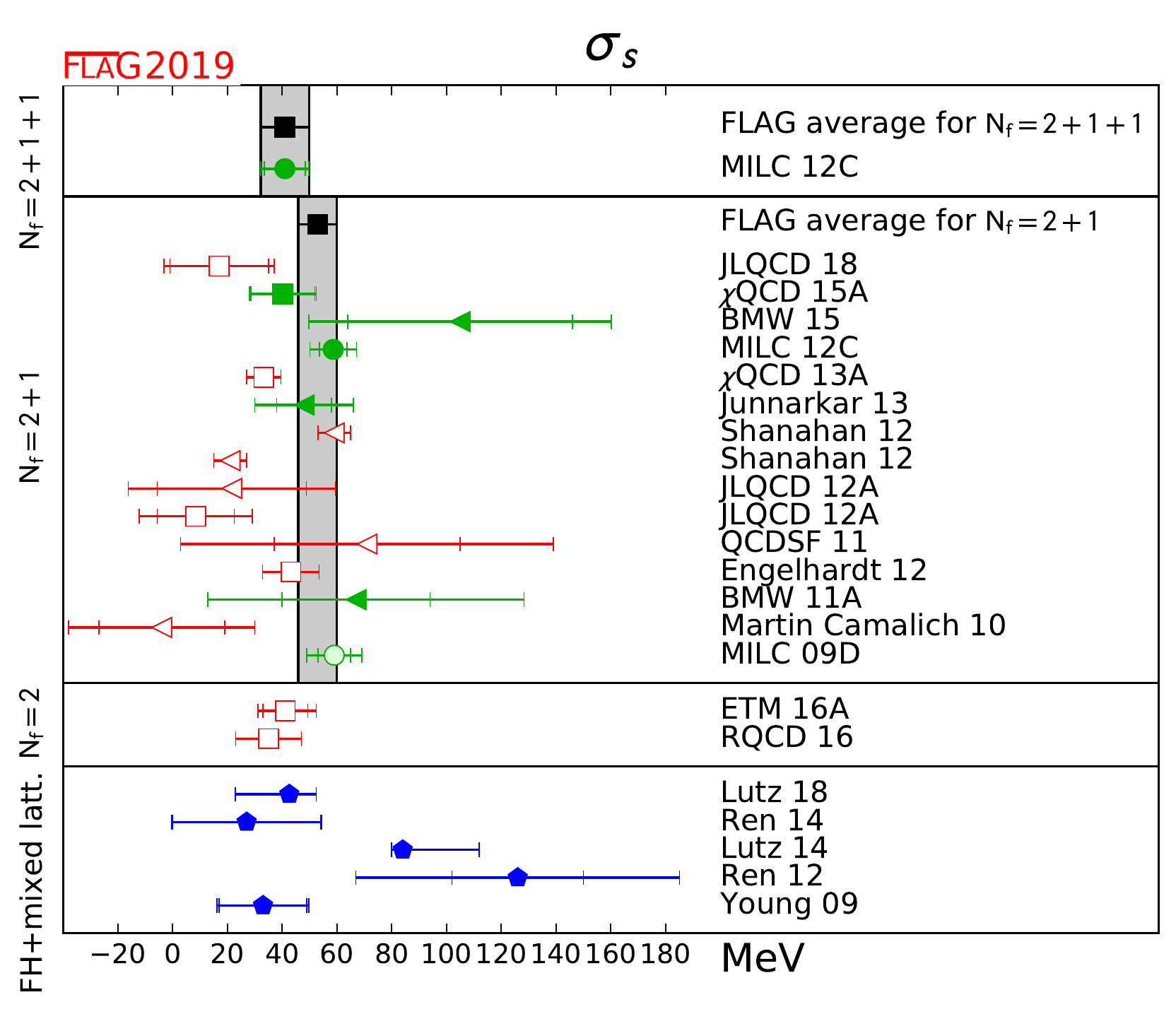}\hss}
\end{minipage}

  \caption{Compilation of the world data on isosinglet \texorpdfstring{\(g_S\)}{g\string_S} presented in the FLAG~\protect\cite{Aoki:2019cca} Review.}
  \label{fig:FLAGsigma}
\end{figure}

The situation of the scalar charges that give the \(\pi N\)-\(\sigma\)
term continues to be unresolved.  As shown in
Fig.~\ref{fig:FLAGsigma}, the \(\sigma_{\pi N}\) results vary between
a cluster around 40, and another around 60, with most controlled
lattice calculations favoring the former. The 2+1+1 flavor average is
the higher value, \(65(13)\), coming from a single-ensemble
calculation by the ETM collaboration~\cite{Alexandrou:2019brg} that includes the charm
quark. The corresponding strange sigma term is known with about 20\%
accuracy, again preferring the smaller value of around 40~MeV.  PNDME~\cite{Park.this}
has presented their first results favoring the smaller value
of \(34(6)\). They also report an additional systematic---a dependence
on the renormalization scheme used. The RI-MOM scheme gives the
smaller value, \(23(7)\) compared to the RI-SMOM. This scheme
dependence needs further attention.

The BMW collaboration~\cite{Varnhorst.this} presented preliminary results for the nucleon
sigma terms for the $u$, $d$, $s$ and $c$ quarks using a stout
staggered action. The values for $\sigma_{qN}$ are consistent with
their earlier results.

\section{Vector Form Factor}

\begin{figure}
\begin{center}
\includegraphics[width=0.45\linewidth]{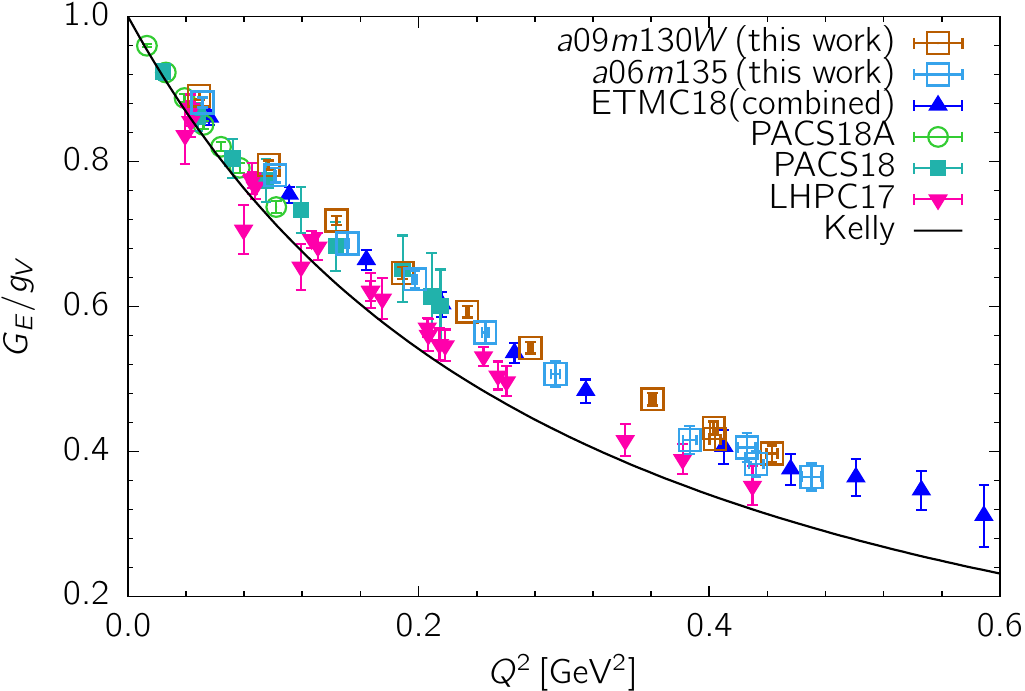}
\includegraphics[width=0.45\linewidth]{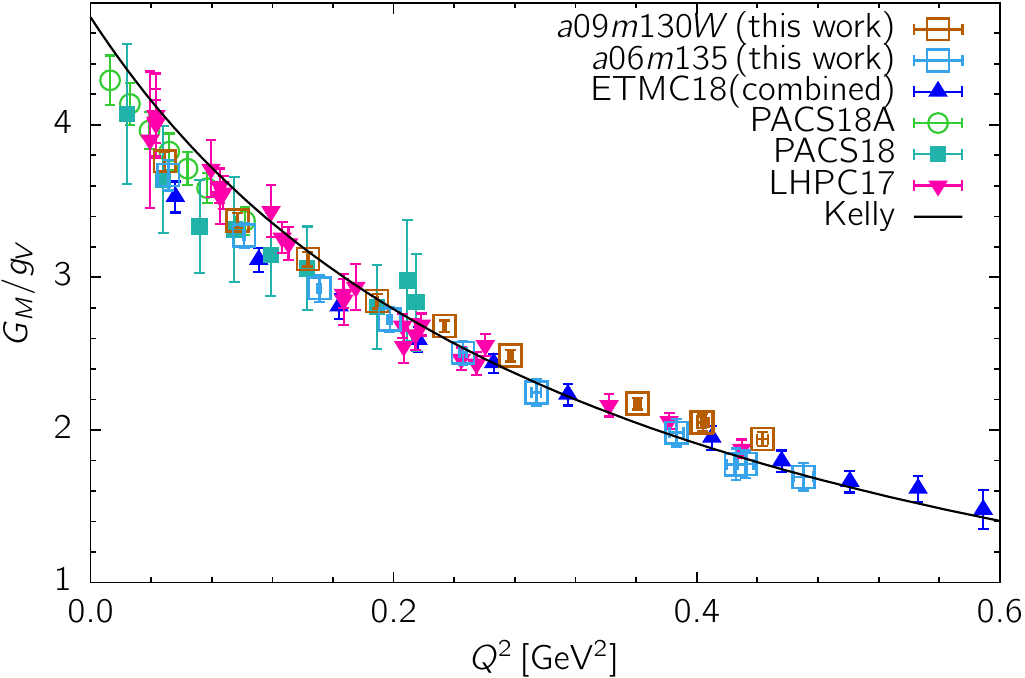}\\
\includegraphics[width=0.45\linewidth]{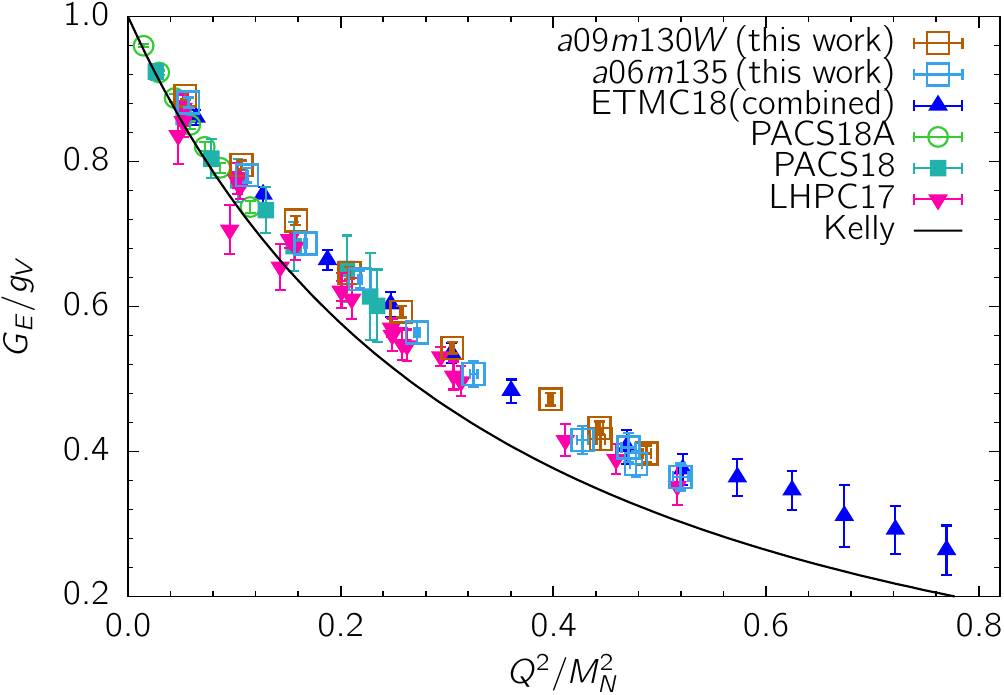}
\includegraphics[width=0.45\linewidth]{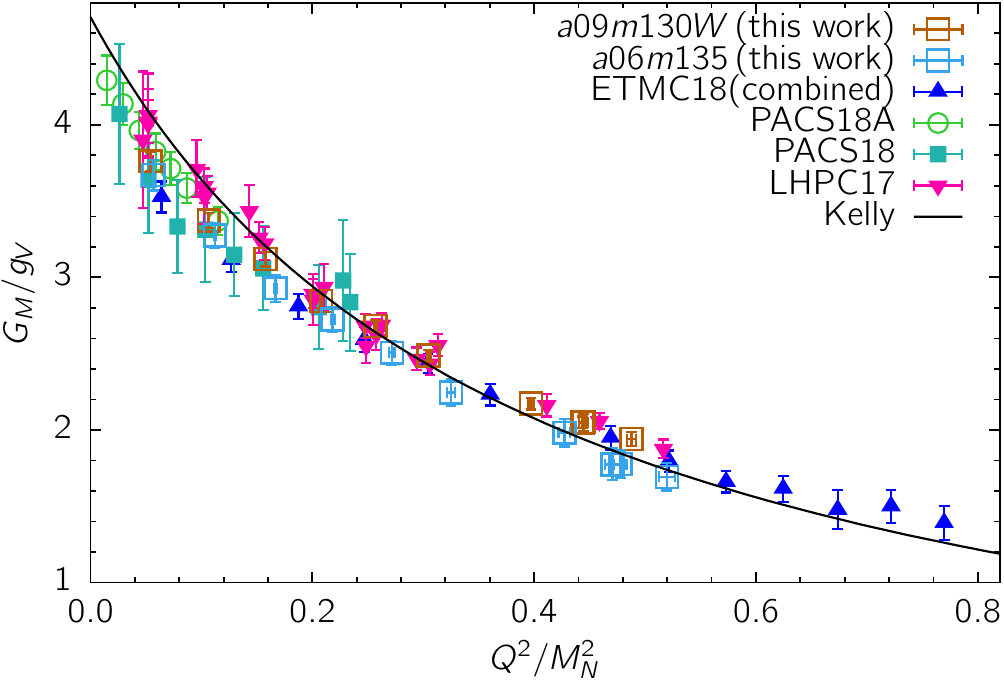}
\end{center}
\caption{World data for the electric and magnetic form factors 
\(G_E\) and \(G_M\) from simulations with $M_\pi < 150$~MeV compiled in 
Ref.~\protect\cite{Jang:2019jkn}. The upper two panels show the data plotted 
versus $Q^2$ and the lower panels versus $Q^2/M_N^2$. The difference is due to 
what quantity is used to set the lattice scale, and is indicative of discretization errors.}
\label{fig:EMformfactors}
\end{figure}

The quantity of highest phenomenological interest for the vector and
axial form factors is their momentum dependence over a range of $Q^2$
values.  The matrix elements of the vector current are, in general,  parameterized in terms
of four form-factors, $F_{1,2,A,3}(Q^2)$.  Of these, \(F_A\) and \(F_3\) are zero when
parity is a good symmetry, and the \(F_1\) and \(F_2\) are usually
combined into the electric and magnetic form-factors \(G_E\)
and \(G_M\) that have a clear interpretation in the Breit frame: 
\begin{eqnarray*}
&\langle N | V_\mu | N \rangle \equiv \bar u \left[ F_1 \gamma_\mu + F_2 \sigma_{\mu\nu}\frac{q_\nu}{2 M}
                                 + F_A \frac {2 i M q_\mu - q^2 \gamma_\mu}{m_N^2} \gamma_5
                                 - i F_3 \sigma_{\mu\nu} \frac{q_\nu}{2 M}\gamma_5  \right] u &\\
&G_E = F_1 - \frac{Q^2}{4 M^2} F_2 \qquad\qquad G_M = F_1 + F_2& \,.
\end{eqnarray*}
The experimental values of these form-factors are very well
parameterized by the Kelly curve, so it was disconcerting that the
lattice data for \(G_E\) (from simulations with $M_\pi < 150$~MeV) lie
systematically above the curve at \(Q^2>0.1~\rm GeV^2\), as shown in
the upper panels in Fig.~\ref{fig:EMformfactors} taken from
Ref.~\cite{Jang:2019jkn}. The \(G_M\) data, on the other hand, tend to
fall below the curve for \(Q^2<0.1~\rm GeV^2\). The large-volume data
from the PACS collaboration~\cite{Shintani:2018ozy}, is an exception,
and tends to lie closer
to the phenomenological results, but are based on very low statistics.  As shown in Tab.~\ref{tab:EMradii},
the isovector magnetic moment \(\mu =
G_M(0)/G_E(0)\equiv \mu^p-\mu^n\), and the electromagnetic
radii, \(\langle r^2_{E,M} \equiv
-6 \left.\frac{d}{dQ^2} \left(\frac{G_{E,M}(Q^2)}{G_{E,M}(0)}\right)\right|_{Q^2=0}\)
determined on the lattice also deviate from their phenomenological
estimates.

The analysis by the PNDME~\cite{Jang:2019jkn} also shows that the lattice
results are much closer to the Kelly determination if the scale is set
from the nucleon mass as shown in the lower two panels in Fig.~\ref{fig:EMformfactors}. The magnitude of the remaining 
disagreement is comparable to the
scale-setting uncertainty, and points to an underestimate of the 
extrapolation systematics as a likely cause.

\begin{figure}
  \begin{center}
    \includegraphics[width=0.49\textwidth]{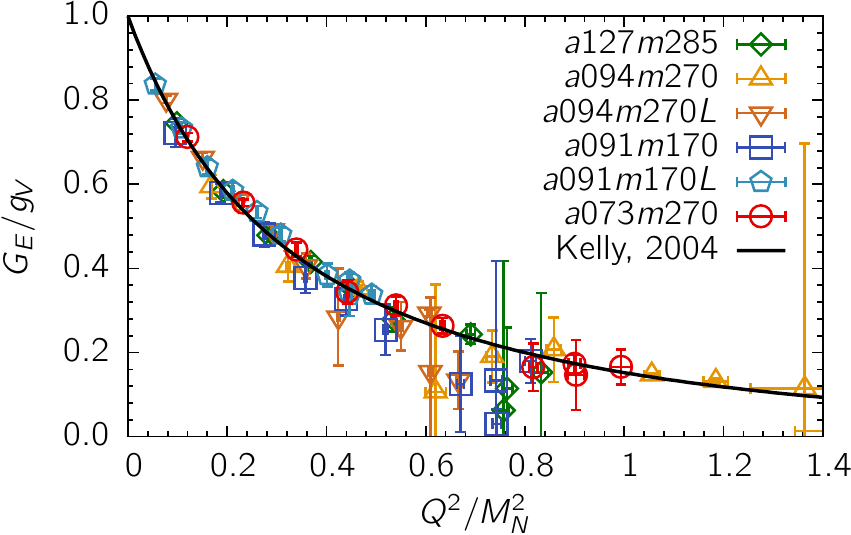}
    \includegraphics[width=0.49\textwidth]{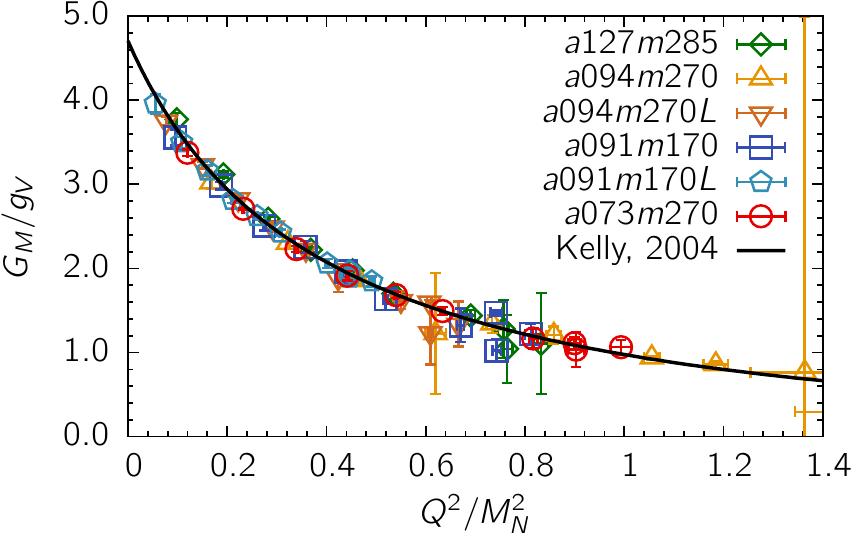}
  \end{center}
  \caption{New results on the electromagnetic form factors from the PNDME collaboration~\protect\cite{Park.this}.}
  \label{fig:GEnew}
\end{figure}

The new results presented by the PNDME collaboration~\cite{Park.this}
using the Clover-on-Clover data (See Fig.~\ref{fig:GEnew}) show much
better agreement with the Kelly curve than the Clover-on-HISQ data
for both $G_E$ and $G_M$. The reason for the improvement with respect to previous results has not been
identified, so further assessment of the systematics needs to be carried out.

\begin{figure}
\begin{center}
  \includegraphics[width=0.85\textwidth]{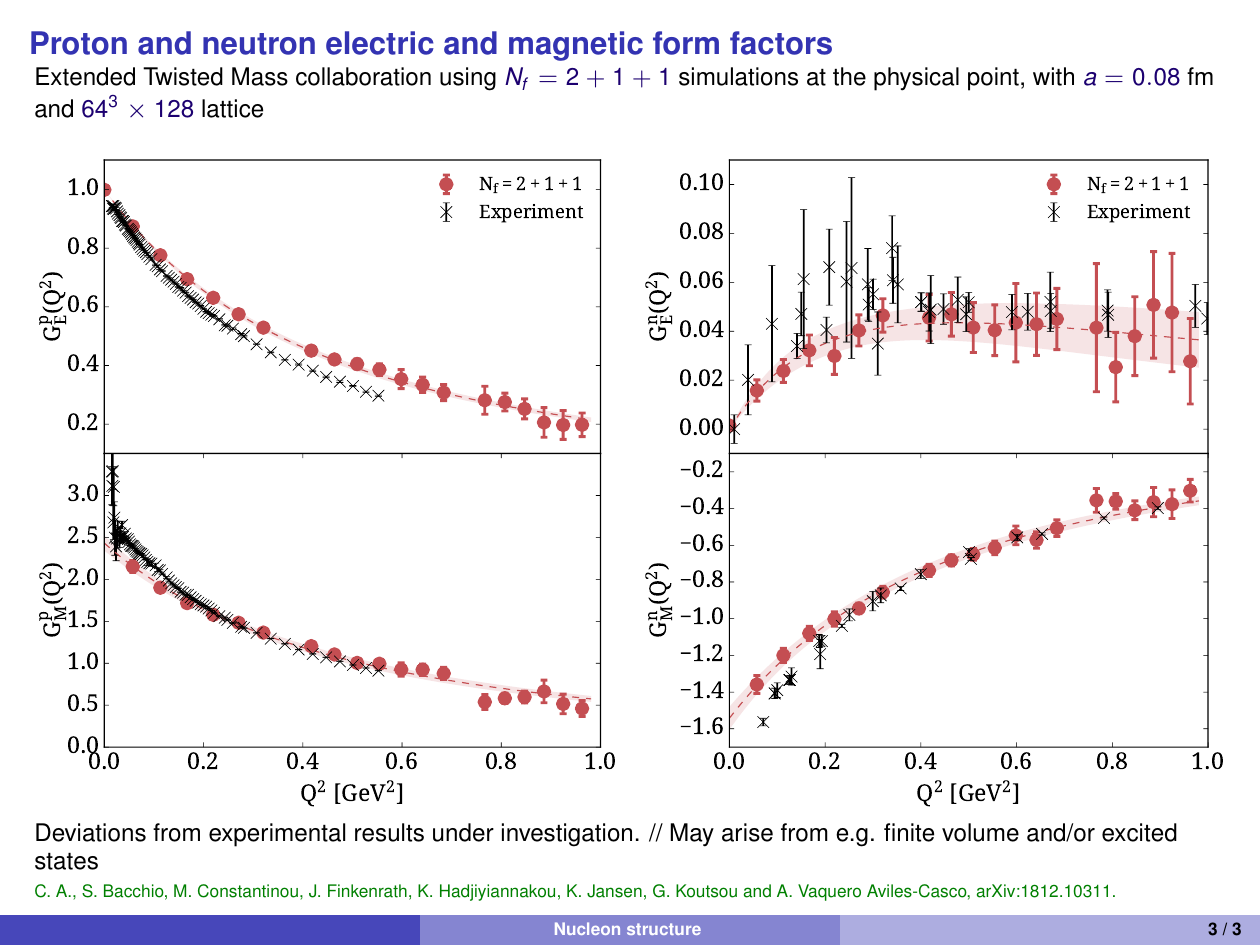}
\end{center}
\caption{Slide from the ETM collaboration showing the results~\protect\cite{Alexandrou:2018sjm} for the electric and magnetic form-factors of the proton including disconnected contributions.}
\label{fig:GEGMDisc}
\end{figure}

\begin{figure}
  \begin{center}
    \includegraphics[width=0.85\textwidth,viewport= 1 20 360 240,clip]{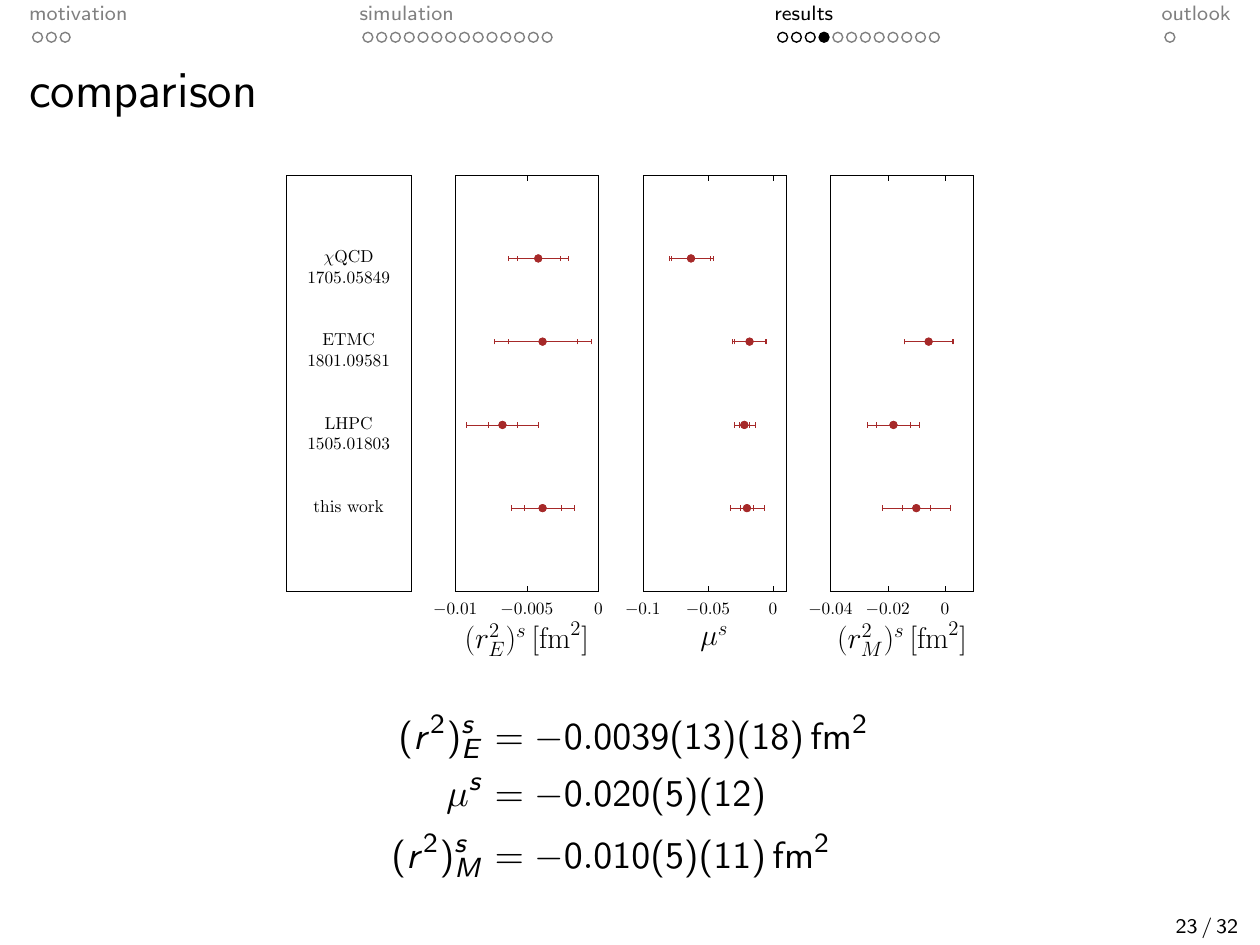}
  \end{center}
\caption{Slide from the Mainz collaboration~\protect\cite{Djukanovic.this} displaying the comparison of their results for the strange contribution to the nucleon with previous results.}
\label{fig:GEGMstrange}
\end{figure}

New results for the disconnected contributions to the form-factors are
available from the ETMC~\cite{Alexandrou:2018sjm} and
Mainz~\cite{Djukanovic.this} collaborations (See
Figs.~\ref{fig:GEGMstrange} and \ref{fig:GEGMDisc}).  The strange
electromagnetic radii and magnetic moment are still determined poorly:
they are within 2 standard errors of zero if statistical and
systematic errors are added in quadrature. More importantly for this
discussion, inclusion of these do not resolve the disagreement with the
experimental results.  Furthermore, the lattice data do not show the
enhancement of the electric form-factor around \(0.2~\rm GeV^2\) as
seen in some experimental data.

\begin{table}
\begin{center}
\begin{tabular}{|l|c|c|c|c|c|c|}
\hline
                     & $M_N$ MeV  & $a$ from           & $Q^2$ Fit  & $r_E$ (fm)   & $r_M$ (fm)   & $\mu$      \\
\hline                                                                                                            
PNDME                  & 953(4)     & $r_1$              & $z^4$      & 0.769(40)& 0.671(90)& 3.94(17)     \\
PNDME a06              & 951(10)    & $r_1$              & $z^4$      & 0.765(14) & 0.704(36)& 3.98(15)     \\
\hline                                                                                                            
LHPC'17     & 912(8)     & $M_\Omega$         & $z^5$      & 0.887(49)    &              & 4.75(15)          \\
\hline                                                                                                            
ETMC'18 & 929(6)     & $r_0$ & dipole     & 0.802(22)  & 0.714(93) & 3.96(16)    \\
ETMC'17 & 941(2)     & $r_0$ & dipole     & 0.808(36)     & 0.732(58)               & 4.02(35)     \\
\hline                                                                                                            
PACS'18  & 958(10)    & $M_\Omega$         & $z^8 | z^7$& 0.915(99)    & 1.437(409)   & 4.81(79)       \\
\hline                                                                                                            
PACS'18A & 942(11)    & $M_\Omega$         & dipole     & 0.875(32) & 0.805(276)  & 4.42(35)        \\
\hline
\end{tabular}
\end{center}
\caption{The values of the neutron magnetic moment, and the electromagnetic charge radii from the various collaborations. Various quoted errors have been added in quadrature. Experimentally,
 \(\mu^p = 1.79\), \(\mu^n = -1.91\), \(r_E = 0.875(6)\) from electrons and \(0.8409(4)\) from muons.}
\label{tab:EMradii}
\end{table}


\section{Axial Form Factor}

The situation for the axial form factors has been much more
questionable as previous results violated the PCAC relation.  The
axial form factors are defined by
\begin{eqnarray}
\langle N | A_\mu | N \rangle &\equiv& \bar u \left[ G_A \gamma_\mu + \tilde G_P \frac{q_\mu}{2 M} \right] \gamma_5 u \\
\langle N | P | N \rangle &\equiv& \bar u  G_P \gamma_5 u
 \end{eqnarray}
From these, the axial radius is extracted using 
\begin{equation}
\langle r_A^2 \rangle \equiv -6 \left. \frac d {dQ^2} \left( \frac {G_A(Q^2)}{G_A(0)}\right) \right|_{Q^2=0}
\end{equation}
Neutrino scattering data pre-2000 were fit to a dipole form to
obtain \(r_A=0.666(17)\) fm and a dipole mass of \(M_A = 1.026(21)\)
GeV~\cite{Bernard:2001rs}, which agrees with those obtained from the
recent analysis of the deuterium system using the $z$-expansion: \(r_A = 0.68(16)\) fm
and \(M_A=1.00(24)\) GeV~\cite{Meyer:2016oeg}. The latter, however,
has much larger uncertainty, which the authors contend is more realistic. 
On the other hand, the
MiniBooNE Collaboration, using the dipole ansatz and a relativistic
Fermi gas model~\cite{Smith:1972xh,Smith:1972xh_erratum}, find that $M_A = 1.35(17)$~GeV
reproduces their double differential cross section for charged current
quasi-elastic neutrino and antineutrino scattering off
carbon~\cite{AguilarArevalo:2010zc}.  Thus, one needs to resolve three
questions: (i) Is the dipole ansatz a good approximation to the $Q^2$
behavior? (ii) If it is, what is the value of $M_A$? And, (iii) 
the cause and resolution of the observation violation of PCAC in the
lattice data discussed below.\looseness-1

\begin{figure}
\begin{center}
\includegraphics[width=0.6\linewidth,viewport=90 125 290 240,clip]{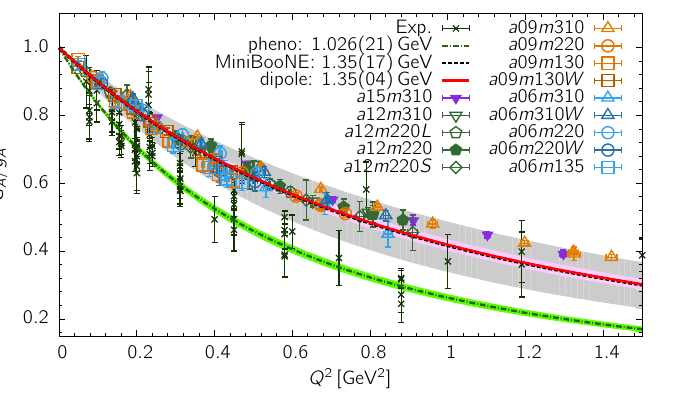}
\end{center}
\caption{Results from the PNDME collaboration on the axial form factors.}
\label{fig:GA}
\end{figure}

The 2017 PNDME lattice results~\cite{Rajan:2017lxk}, shown in Figure~\ref{fig:GA}, lie
closer to the data from the MiniBooNE experiment.  In fact a dipole
fit gives \(r_A=0.505(13)(6)\) fm and \(M_A=1.35(3)(2)\) GeV.  A
$z$-expansion fit gives an even smaller slope at the
origin: \(r_A=0.481(58)(62)\) fm and \(M_A=1.42(17)(18)\) GeV. Their
main conclusion was that the data show little variation versus the
lattice spacing $a$ and the pion mass $M_\pi$, however, to extract
$r_A$, more data near \(Q^2=0\) are needed. Results from the ETMC collaboration 
were similar and gave \(M_A=1.322(42)(17)\)~GeV~\cite{Alexandrou:2017hac}. 

New results on the disconnected contribution, in particular the
strange contribution, were also presented this year by the Mainz
collaboration~\cite{Djukanovic.this}.

\section{PCAC Relation}

\begin{figure}
  \begin{center}
    \includegraphics[page=1,viewport=0 120 175 250,clip,width=0.7\textwidth]{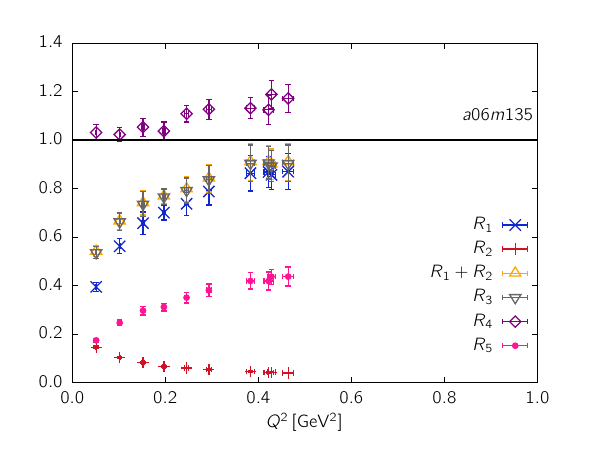}
  \end{center}
\caption{Various ratios of extracted form factors describing the violation of PCAC.}
\label{fig:PCACprob}
\end{figure}

A major advance this year was a credible resolution to the violation
of the PCAC relation on the lattice~\cite{Jang:2019vkm}.  The
situation previously could be summarized as: even though the Axial
Ward Identity: \(\partial_\mu A^\mu = 2 \hat m P\) was indeed
satisfied by the correlation functions on the lattice, the
corresponding relation between the extracted form-factor written as 
\(\left(R_1 \equiv \frac{2 \hat m G_P}{2 M G_A} {2 M
G_A}\right) + \left(R_2\equiv\frac{Q^2\tilde G_P}{4 M^2 G_A}\right)
=1 \) was strongly violated. These violations grew as $M_\pi \to 0$
and $Q^2 \to 0$.  In addition, the related hypothesis of pion-pole
dominance, \(R_4 \equiv \frac{ 4\hat m M_N G_P}{M_\pi^2 \tilde G_P} =
1\) was equally badly violated~\cite{Rajan:2017lxk}. This is illustrated in
Fig.~\ref{fig:PCACprob}, where we also plot the
combination \(R_3 \equiv \frac{(Q^2+M_\pi^2)\tilde G_P}{4 M_N^2 G_A}
=1 \), and \(R_5\), the expected \(O(a)\) correction to \(R_1\)
calculated with the the unimproved axial current.

Bali {\it et al.}~\cite{Bali:2018qus} proposed a fix for this using
projected currents \(A_\mu^\perp = (g_{\mu\nu} - \bar p_\mu\bar
p_\nu/\bar p^2) A_\nu\), with 4-momentum \(\bar p\) the reflection
of \(p\) about the light cone. The diagonal matrix element of the
added term should be zero in any spin-half state, so this current
should provide an equally good determination of the axial form-factors
of the neutron, but possibly removing excited state effect due to
transition matrix elements.  They then suggested shifting the pseudoscalar
operator by a similar term, \(P^\perp = P - (\frac{1}{2im_q} \bar
p_\mu\bar p_\nu/\bar p^2) \partial_\mu A_\nu \) but enforcing \(p^\mu
A_\mu^\perp = 2 \hat m P^\perp\).  The matrix element of $P^\perp$ was
much larger and helped satisfy PCAC. Since the quark masses $m_q$ are
small numbers, it was {\it a priori} unclear whether the large shift
could be caused merely by \(O(a^2)\) effects being amplified by the
smallness of the quark mass. Also, it was not clear why this
projection was needed in the first place.  Lastly, this construct did
not fix the small value of $g_P^\ast$. An interesting feature of this proposed solution is that the
projection almost completely eliminates contributions from the \(A_4\), 
and its large excited state effect, instead 
$A_4^\perp$ is dominated by contributions from $A_i$. 

\begin{figure}
  \begin{center}
    \includegraphics[page=2,width=0.85\textwidth,viewport=9 10 357 263, clip]{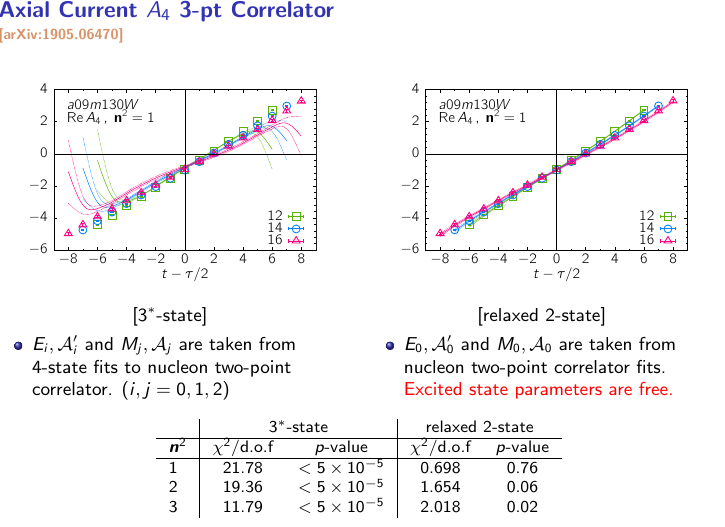}
  \end{center}
\caption{Slide from PNDME~\protect\cite{Jang.this} showing fits to the \(A_4\) correlation function showing that the mass-gap determined from the two-point function does not fit the \(A_4\) correlator (top left), but leaving it free gives a good fit (top right). The bottom table provides the \(\chi^2\) values for the different fits.}
\label{fig:PCACA4corr}
\end{figure}

\begin{figure}
  \begin{center}
    \includegraphics[page=3,width=0.85\textwidth,viewport=9 100 357 263, clip]{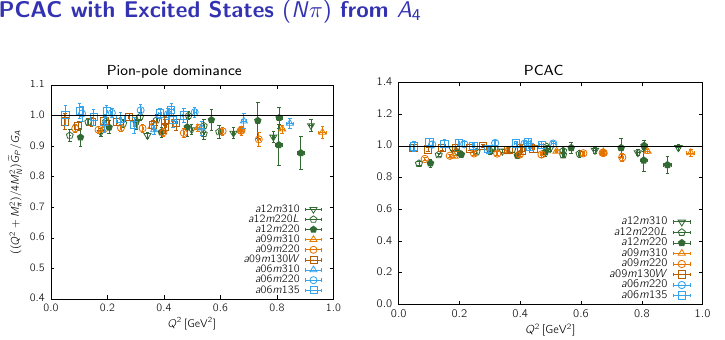}
  \end{center}
\caption{Slide from PNDME~\protect\cite{Jang.this} claiming  solution to the PCAC puzzle in terms of a low-lying excited state that had been missed.  Both the ratios measuring violation of pion-pole dominance (left) and of PCAC relation between form factors (right) are unity if the low-lying excited state is accounted for.}
\label{fig:PCACsoln}
\end{figure}

The PNDME collaboration, on the other hand showed~\cite{Jang.this} that the problem is
resolved by including low-mass excited states when removing the
excited state contamination, and the matrix element of the \(A_4\)
current is key to their solution.  In particular, when the mass-gap
between the first-excited and the ground-state is extracted from the
two-point function, the matrix elements of the spatial
components \(A_i\) are well-fit, but the matrix elements of \(A_4\)
has very large $\chi^2$.  These latter matrix elements can, however, be
fit if the mass-gap is left free (see Fig.~\ref{fig:PCACA4corr}). 
The fit value of the mass-gap is close to 
a non-interacting intermediate \(N\pi\) state. It is much smaller but 
cannot be ruled out by fits to the two-point fit on the basis of $\chi^2$.  
The \(A_i\) correlators, on
the other hand, fit equally well with any of these mass-gaps, but the
extrapolated results turn out to be different.  What PNDME finds (see
Fig.~\ref{fig:PCACsoln}) is that using the much smaller mass-gap, the
PCAC relation, as well as the pion pole-dominance ($g_P^\ast$), are satisfied up
to small corrections that are of size that can plausibly be attributed to \(O(a)\)
effects.\looseness-1

\begin{figure}
  \begin{center}
    \includegraphics[width=0.32\textwidth]{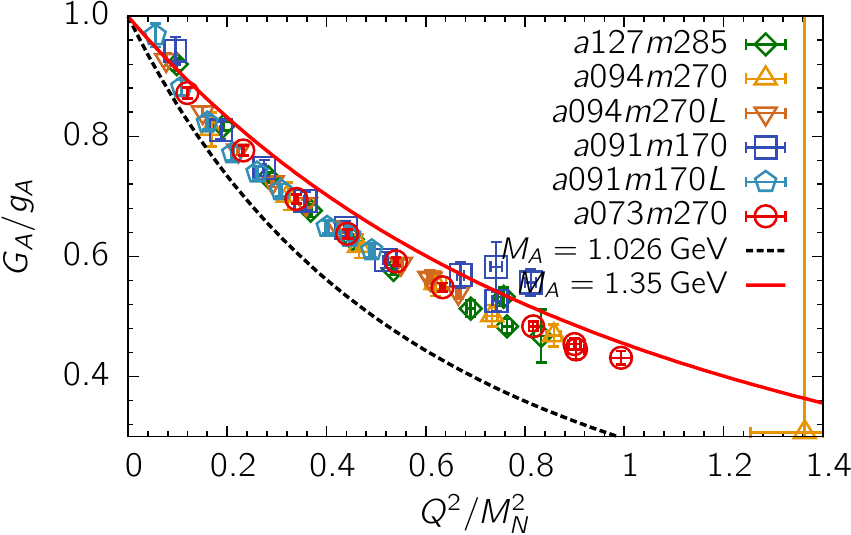}
    \includegraphics[width=0.32\textwidth]{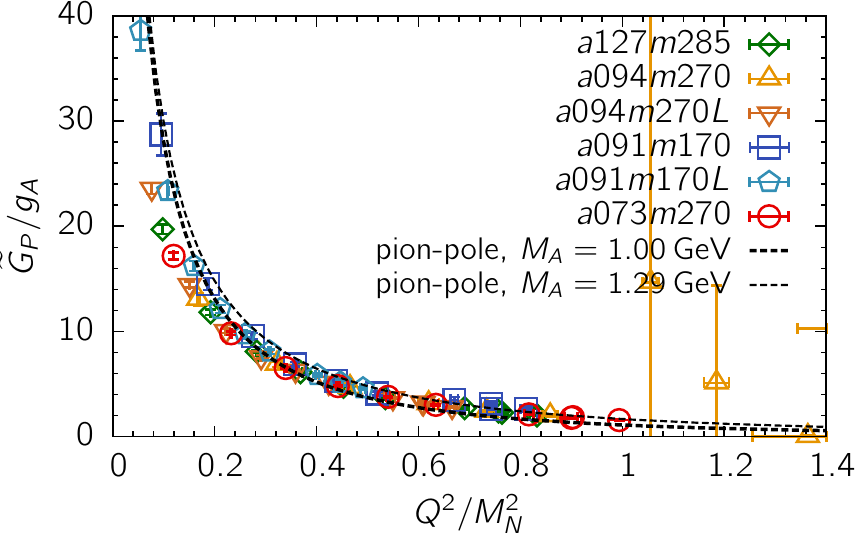}
    \includegraphics[width=0.32\textwidth]{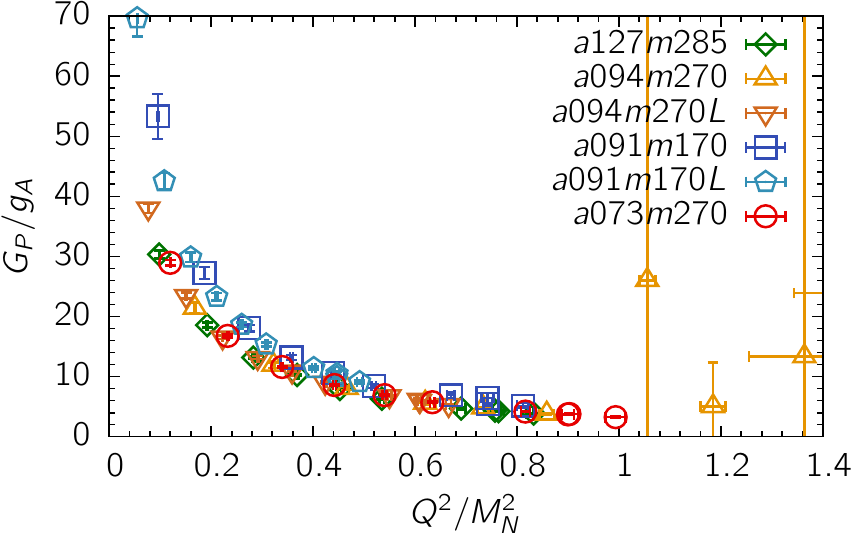}
  \end{center}
  \caption{Isovector form-factors \(G_A\), \(\tilde G_P\) and \(G_P\) obtained by the PNDME collaboration~\protect\cite{Park.this} taking into account the low-lying \(N\pi\) excited state contamination.}
  \label{fig:GAnew}
\end{figure}

The net outcome is a new determination of the form factors \(G_A\), \(G_P\) and \(\tilde G_P\).  They present results for both the Clover-on-HISQ~\cite{Jang.this} and Clover-on-Clover~\cite{Park.this} analysis.  The results of the two analysis are similar: the latter results are reproduced here in Fig.~\ref{fig:GAnew}.

\section{Other matrix elements}

In this section, we briefly comment on the recent advances in
calculating the nucleon matrix elements of a few other operators.
These calculations are all preliminary---often the first calculations
of their nature on the lattice---and the systematics 
are yet to be controlled.\looseness-1

\subsection{CP violation from BSM}
\begin{figure}
\includegraphics[width=0.45\linewidth]{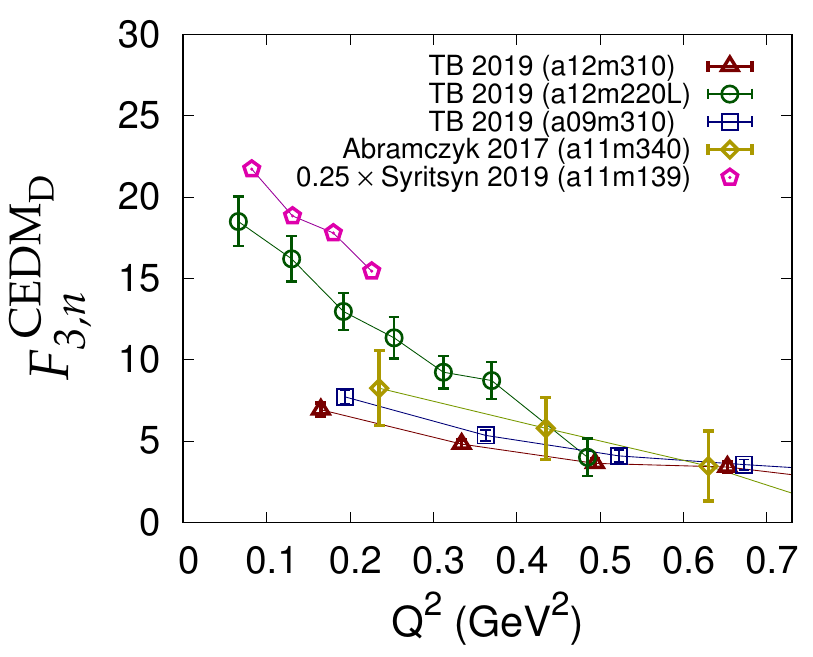}
\includegraphics[width=0.45\linewidth]{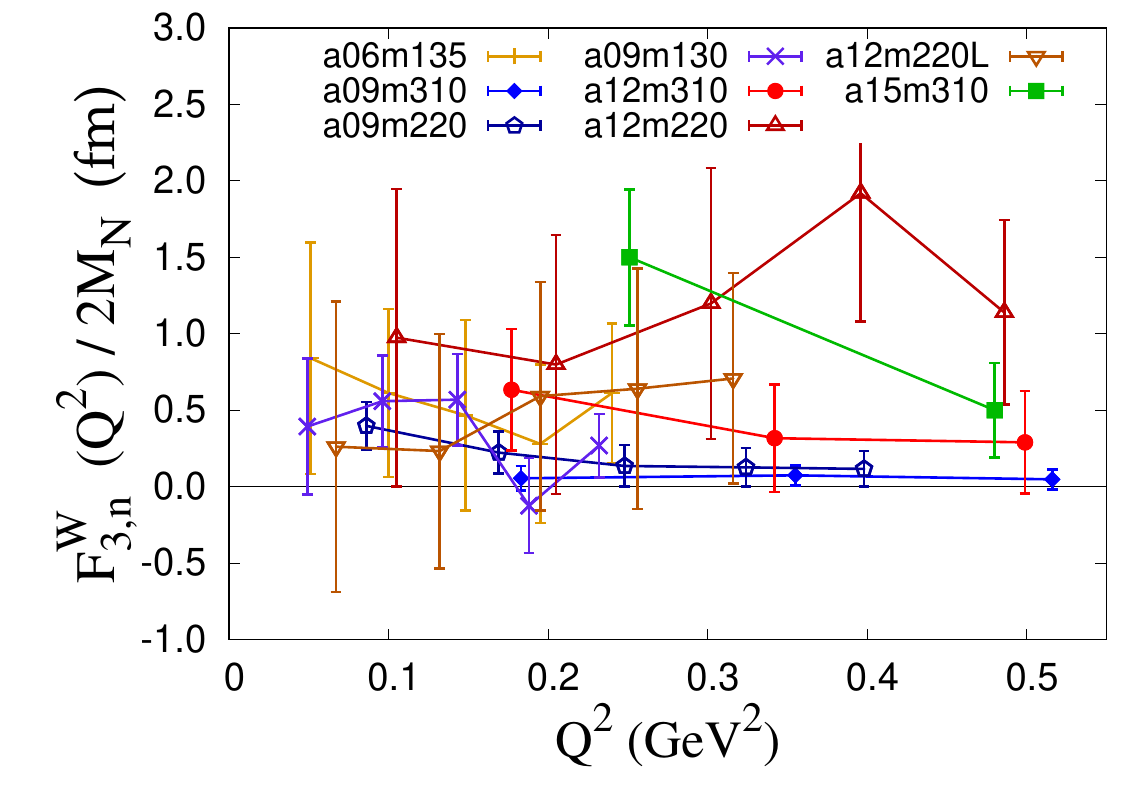}\\
\includegraphics[width=0.45\linewidth]{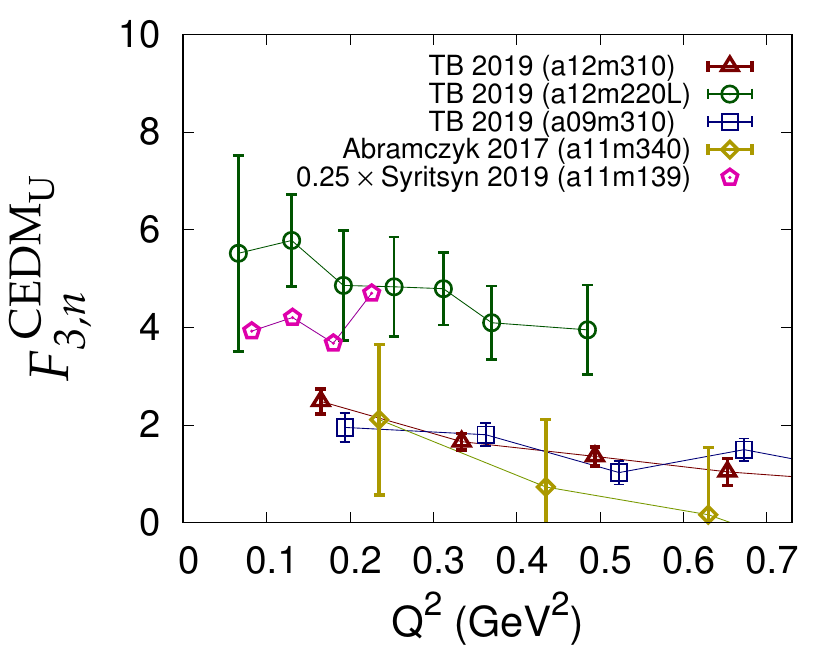}
\includegraphics[width=0.45\linewidth]{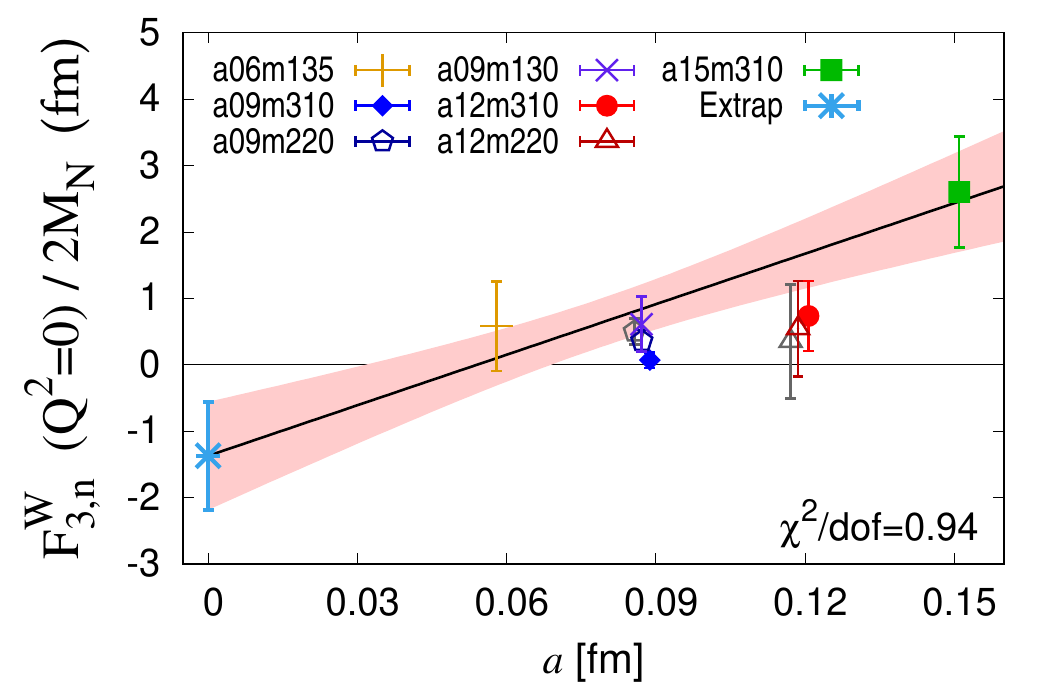}
\caption{Comparison of the extraction of the electric dipole moment of the nucleon from the chromo-EDM and the Weinberg three-gluon operators, taken from the talk by B.~Yoon~\protect\cite{Yoon.this}. Only connected diagrams have been calculated, and the data shown are the lattice numbers without any renormalization.}
\label{fig:nEDM}
\end{figure}

The calculation of nEDM due to the \(\Theta\)-term is discussed in the plenary lecture by H.~Ohki~\cite{Ohki.this}. Last year also saw a lot of activity in the calculation of the neutron EDM arising from the operators beyond the standard model, especially those due to the quark chromo-EDM and the Weinberg operator.  In a parallel talk, Boram Yoon~\cite{Yoon.this}, summarized the unrenormalized results for the \(F_3\) form factors, which are reproduced in Fig.~\ref{fig:nEDM}. As can be seen, the statistical errors in the case of the Weinberg operator are not yet under control.  For the quark chromo-EDM, the data from different groups using different actions, naively show a very large quark-mass dependence. Further analysis is needed to understand whether the renormalization and operator mixing effects can explain this.

\subsection{Gravitational Moments}

\begin{figure}
  \begin{center}
    \setlength{\tabcolsep}{0.5pt}
    \begin{tabular}{ccc}
\includegraphics[width=0.33\linewidth]{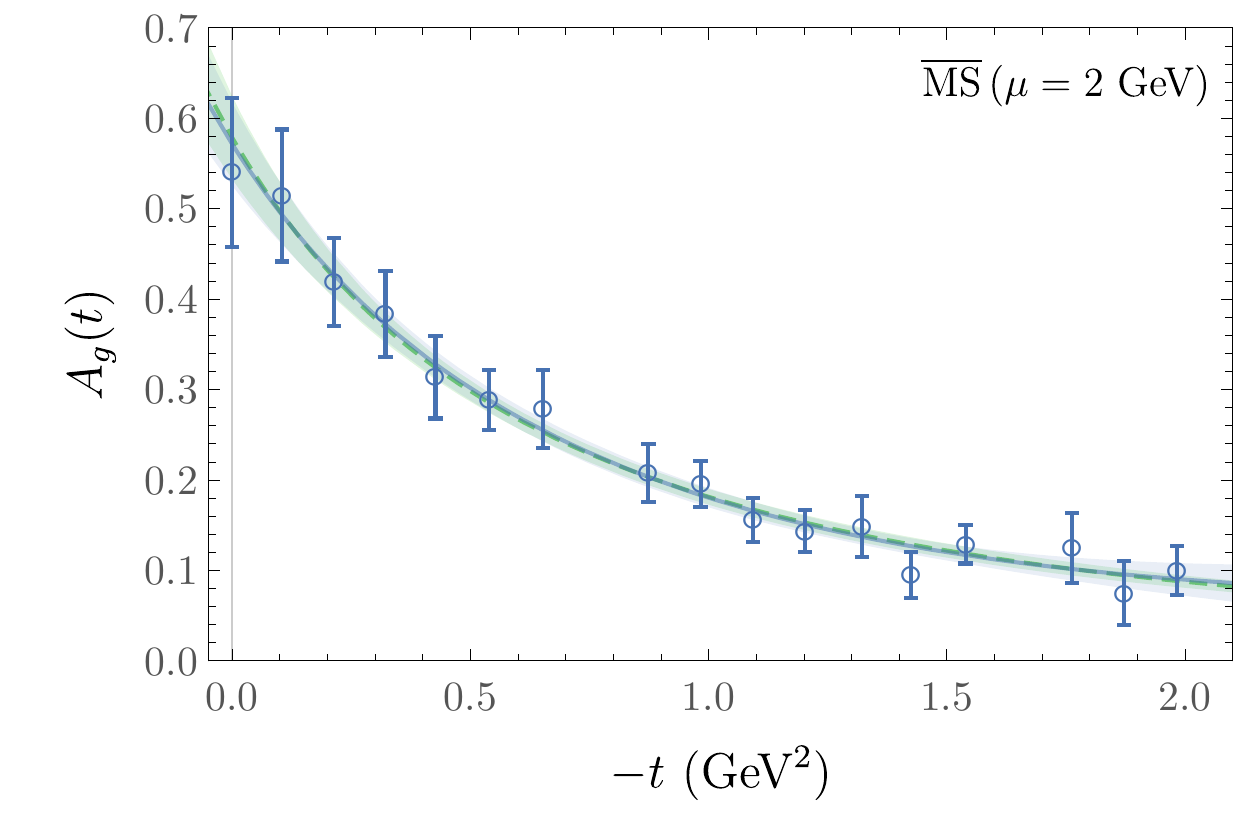}&
\includegraphics[width=0.33\linewidth]{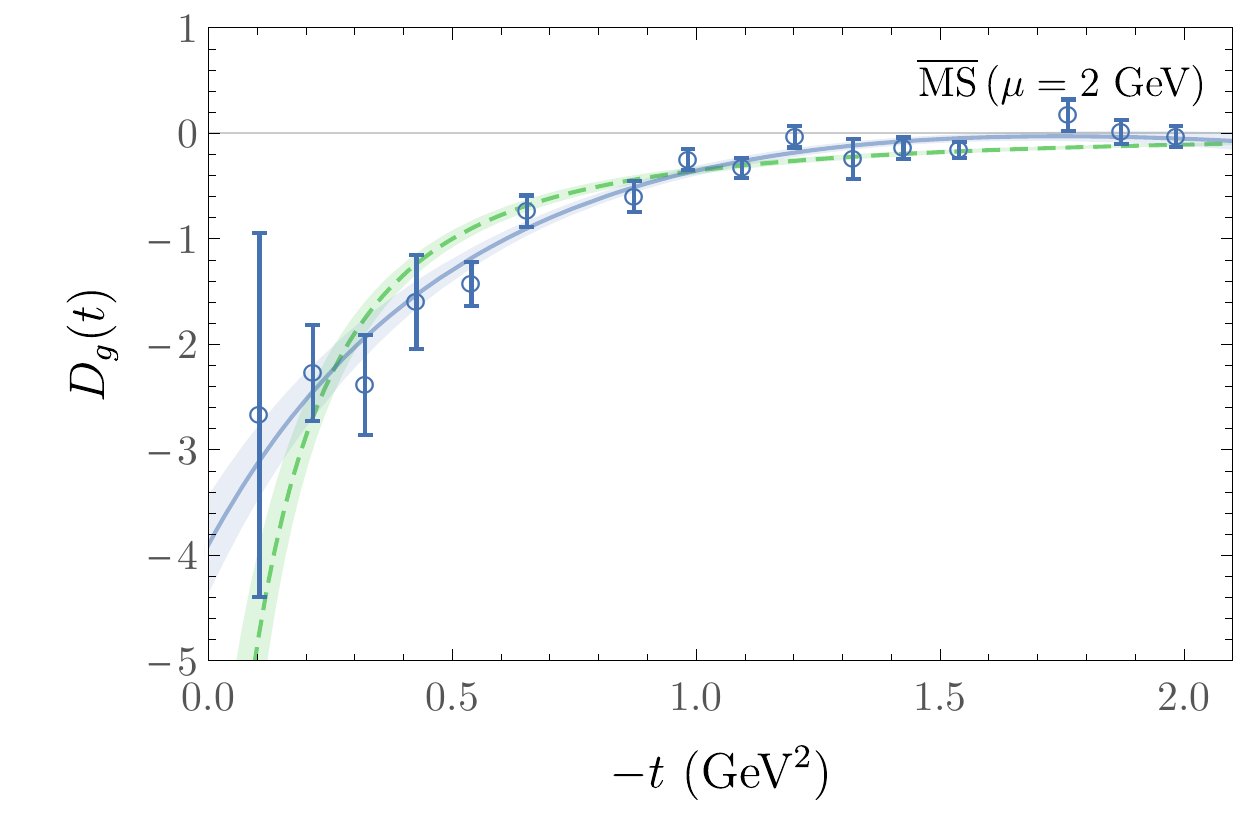}&
\includegraphics[width=0.33\linewidth]{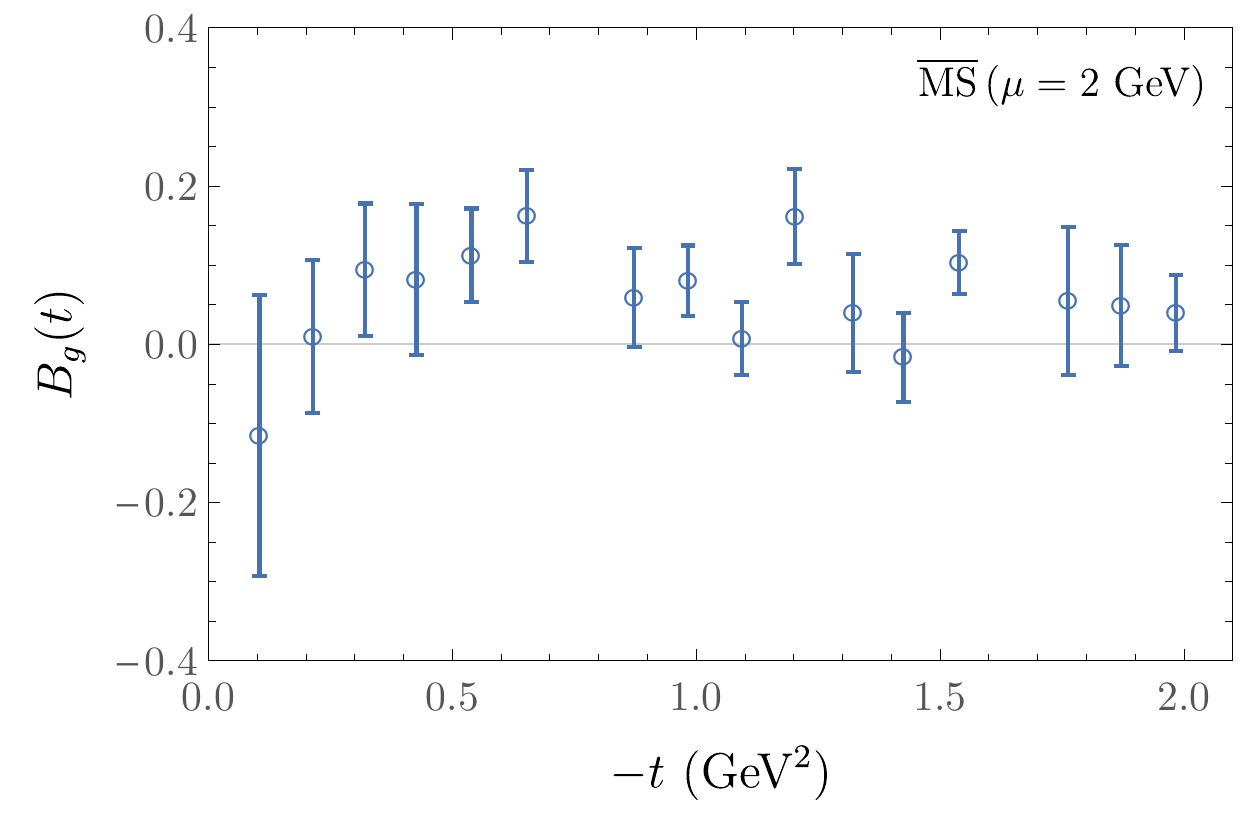}\\
\includegraphics[width=0.33\linewidth]{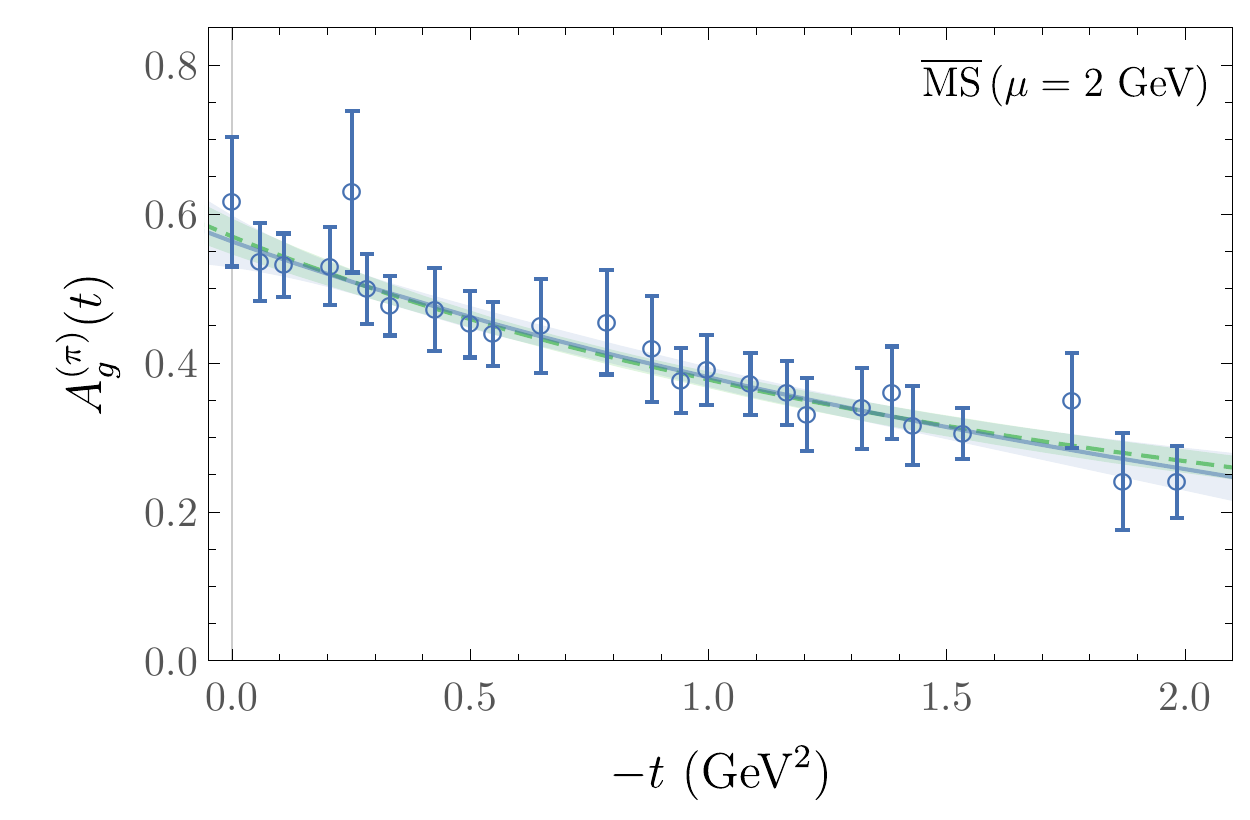}&
\includegraphics[width=0.33\linewidth]{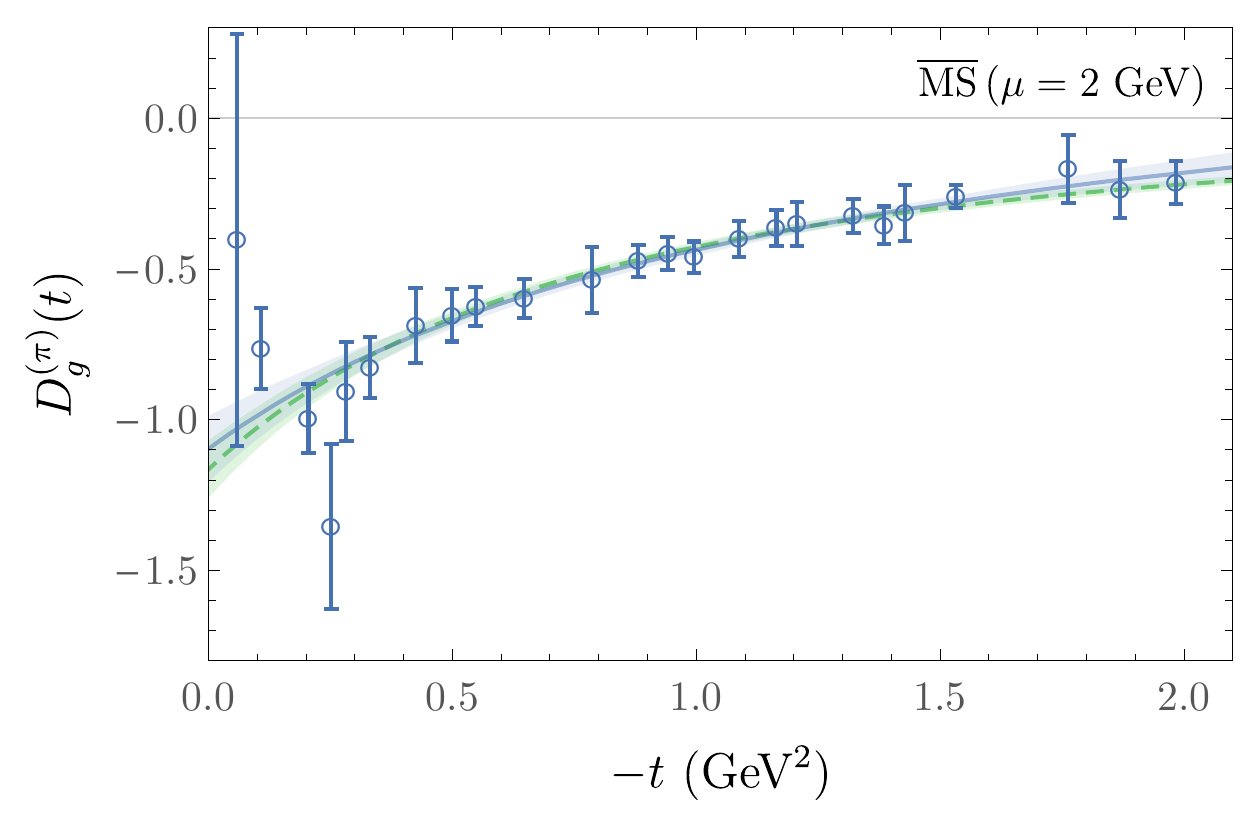}
    \end{tabular}
  \end{center}
\caption{The gravitational moments of the nucleon by Shanahan and Detmold~\protect\cite{Shanahan:2018pib}.}
\label{fig:gravmom}
\end{figure}

\begin{figure}
  \begin{center}
    \includegraphics[width=0.75\textwidth]{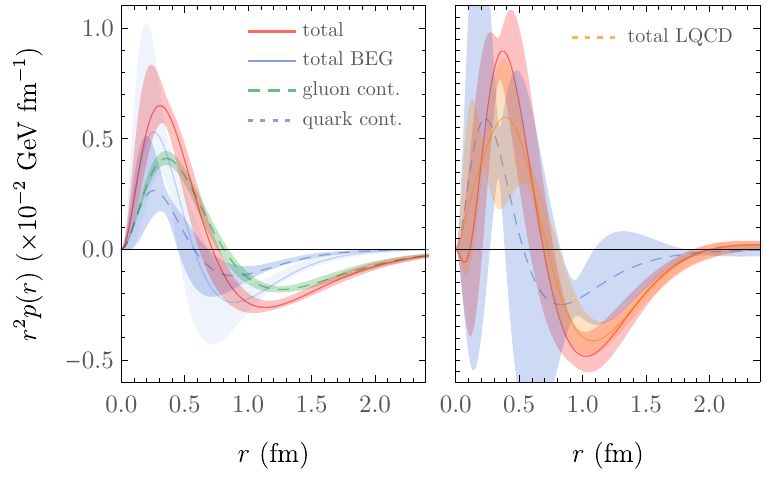}\\
    \includegraphics[width=0.75\textwidth]{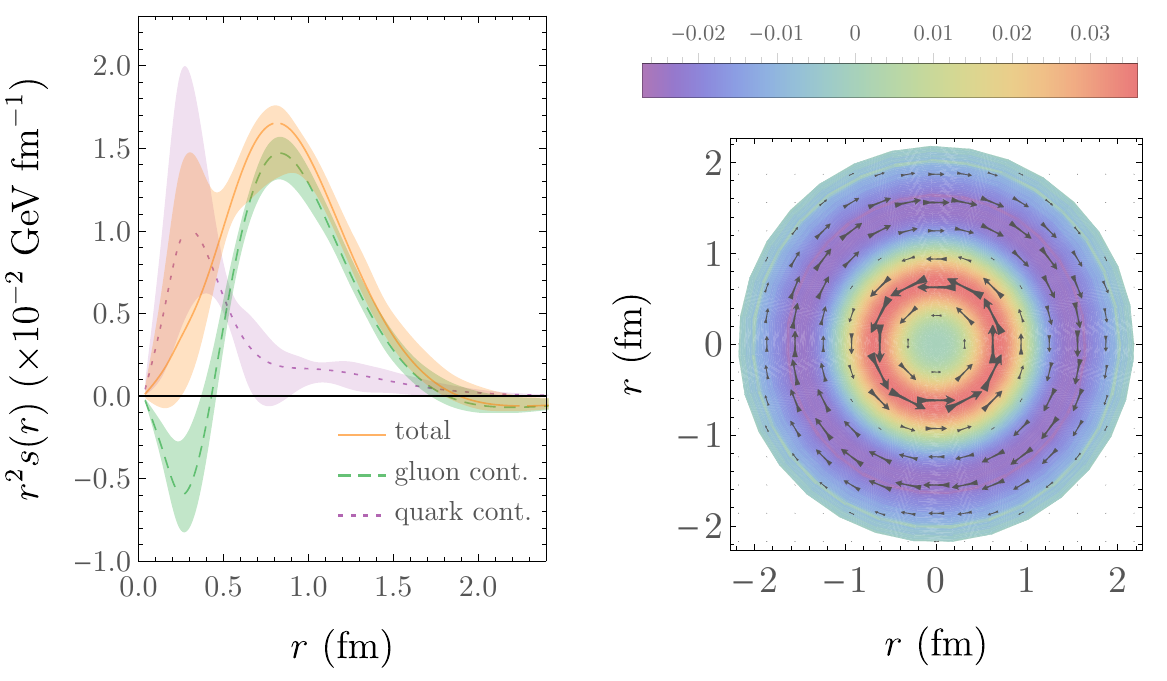}
  \end{center}
\caption{Pressure and shear distribution inside the nucleon by Shanahan and Detmold~\protect\cite{Shanahan:2018nnv}.}
\label{fig:pressure}
\end{figure}

In the last year, Shanahan and Detmold have provided the first calculations~\cite{Shanahan:2018pib,Shanahan:2018nnv} of the gravitational moment of the nucleon (Fig.~\ref{fig:gravmom}) and the pressure and shear distribution within the nucleon (Fig.~\ref{fig:pressure}).  These calculations have been done with a large pion mass ensemble, but provide the first estimate of the mechanical radius of the proton as \(0.71(1)\)~fm.  Their data for the pressure and shear distributions, when fit using model-independent $z$-fits, have large errors, and the systematics need further study.

\subsection{Magnetic Polarizability}

\begin{figure}
  \begin{center}
    \includegraphics[width=0.4\textwidth]{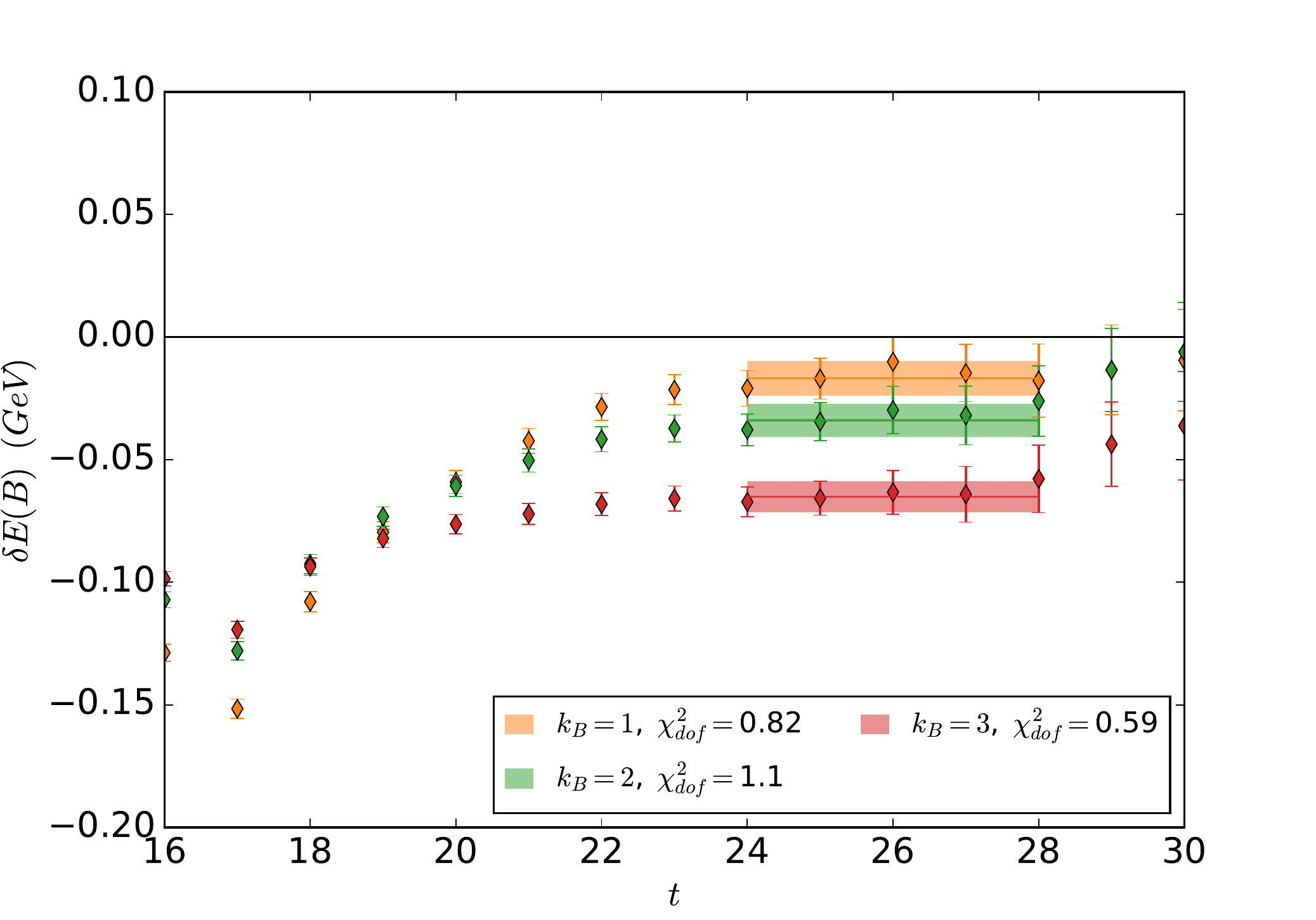}
    \includegraphics[width=0.4\textwidth]{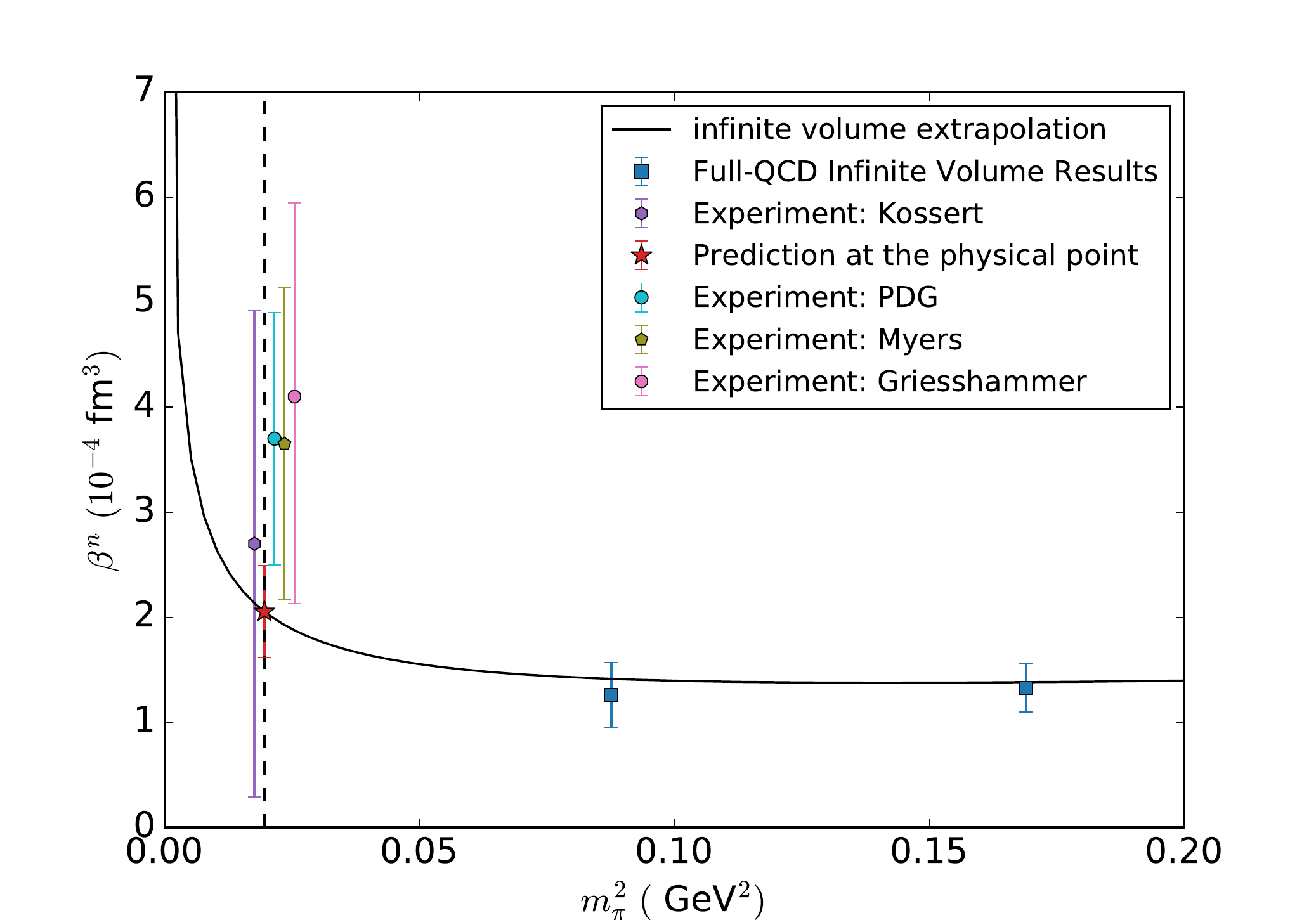}
  \end{center}
\caption{Magnetic polarizability of the nucleon~\protect\cite{Bignell:2018acn}.}
\label{fig:magnpol}
\end{figure}

First results on calculating the polarizability of the nucleon were presented by the Adelaide collaboration~\cite{Bignell:2018acn}, and are reproduced here as Fig.~\ref{fig:magnpol}. These first calculations need further study before the systematics, especially the chiral extrapolation, is understood.

\section{Future}

The lessons from these calculations is that though lattice calculations have matured, we still need to improve our approach to understanding systematic errors.  In particular, it is easy to inadvertently increase the systematic errors in our quest to reduce systematic errors.  In particular, the question that we need to ask is not whether any systematic is visible in a given data set, but, rather, how \emph{large} a systematic could be masked in it.  The approach advocated by a number of groups that use model averaging is a step in the right direction, but does not solve the problem since the space of models averaged may not be large enough, and the weights assigned to the conservative models may be too low.  Partial solution would be to move over to properly chosen blind analysis techniques, since, at least, unconscious bias gets minimized in such approaches.

There are still many operators, calculating whose  matrix elements needs more work. We need a better control (statistical and systematic) of disconnected diagrams, of scalar matrix elements, as well as matrix elements of higher dimensional (BSM) operators.

\section{Acknowledgments}

Thanks to Yong-chull Jang for help with the presentation. I would also like to thank everybody that sent me unpublished materials or drew my attention to their work. For the data from our own collaborations, I would like to thank OLCF, NERSC, USQCD  for computer allocations allowing these calculations, MILC Collaboration for HISQ lattices, and LANL LDRD and DOE office of science for supporting this research and me.


%
\hfuzz=3pt
\printbibliography

\end{document}